\documentstyle[12pt]{article}
\begin{document}
\begin{center}

\null
\vskip-1truecm
\rightline{IC/95/6}
\vskip1truecm
{International Atomic Energy Agency\\
and\\
United Nations Educational Scientific and Cultural Organization\\
\medskip
INTERNATIONAL CENTRE FOR THEORETICAL PHYSICS\\}
\vskip2.5truecm
{\bf CHIRAL FERMIONS ON THE LATTICE}
\vskip2truecm
S. Randjbar--Daemi\quad and\quad J. Strathdee\\
International Centre for Theoretical Physics, Trieste, Italy.\\
\end{center}
\vskip1truecm

\vskip8truecm
\begin{center}
{MIRAMARE -- TRIESTE\\
\medskip
January 1995\\}
\end{center}

\newpage

\baselineskip=24pt

\centerline{ABSTRACT}
\bigskip

The overlap approach to chiral gauge theories on arbitrary $D$--dimensional
lattices is studied. The doubling problem and its relation to chiral anomalies
for $D=2$ and 4 is examined. In each case it is shown that the doublers
can be eliminated and the well known perturbative results for chiral anomalies
can be recovered. We also consider the multi--flavour case and give the general
criteria for the construction of anomaly free chiral gauge theories on
arbitrary
lattices. We calculate the second order terms in a continuum approximation to
the overlap formula in $D$ dimensions and show that they coincide with the
bilinear part of the effective action of $D$--dimensional Weyl fermions coupled
to a background gauge field. Finally, using the same formalism we reproduce the
correct Lorentz, diffeomorphism and gauge anomalies in the coupling of a Weyl
fermion to 2--dimensional gravitational and Maxwell fields.

\newpage

\section{Introduction}

A problem of long standing in particle physics has been to invent a lattice
regularization scheme for chiral fermions. A solution would be desirable
because it could be employed in conjunction with the well developed methods of
lattice gauge theory to study non--perturbative phenomena such as chiral
symmetry breaking in the standard model. Some progress seems to have been made
recently. Narayanan and Neuberger [1], influenced by earlier work of Kaplan
[2] and others [3], have constructed a lattice representation for the vacuum
amplitude of chiral fermions coupled to a background gauge field that seems to
meet all the physical requirements. In particular, it has been shown that in
the continuum framework it carries the expected chiral anomalies in two [4] and
four [5] dimensions. A lattice analysis of the 2--dimensional chiral Schwinger
model has also been performed [6]. Whether this representation will turn out to
be useful for numerical studies is a separate question.

The proposal of Narayanan and Neuberger is to embed the $D$--dimensional
Euclidean chiral theory in a $D+1$--dimensional auxiliary system involving
twice as many fields. For example, each 2--component Weyl spinor in 4
dimensions is represented by a 4--component Dirac spinor in $4+1$--dimensions.
The embedding theory involves a mass parameter, $\Lambda$, which is eventually
taken to infinity in such a way that all the physically irrelevant auxiliary
states are projected out. The surviving states, zero modes, so to speak, of the
embedding theory, represent chiral fermions in $D$--dimensions. The mechanism
is somewhat obscure, to us at least, but it seems to work. It may prove useful
and, therefore, we believe that it deserves to be better understood. Our aim in
this paper is to describe the mechanism as we understand it and to present some
calculations that we hope will clarify its function.

By way of historical introduction we begin with a brief description of the
chirality problem for fermions on a lattice. In 4--dimensional Euclidean
spacetime the Green's function for a 2--component Weyl spinor is given, in the
continuum, by

\newpage

\setcounter{equation}{0}
\renewcommand{\theequation}{1.\arabic{equation}}
\begin{eqnarray}
G(k)^{-1} & = & k_\mu\ \sigma_\mu\nonumber \\
& = & ik_4+\underline{k}\cdot\underline{\sigma}
\end{eqnarray}
where the momenta, $k_\mu$, take values on the Euclidean 4--plane. Physical
states are associated with poles at $k_4=\pm i\vert\underline{k}\vert$. To put
this on a lattice one must replace $k_\mu$ by a vector function $C_\mu (k)$
defined over a torus. In effect, the momenta must be viewed as angular
variables,
$k_\mu\sim k_\mu +2\pi /a$ where $a$ represents the lattice spacing. Physically
meaningful states must have $k_\mu a\ll 1$ and the Green's function should
reduce to (1.1) near $k_\mu a=0$, i.e. $C_\mu (k)\sim k_\mu$.

There is a topological theorem due to Poincar\'e and Hopf [7] according to
which
the zeroes of a vector function on a torus do not occur singly. The occurrence
of zeroes is conditioned by the requirement
\begin{equation}
\sum_{zeroes}\ {\det C\over \vert\det C\vert} =0
\end{equation}
where $\det C=\det_{\mu\nu}(\partial C_\mu /\partial k_\nu )$, evaluated at
$C_\mu =0$. Since physics requires $C_\mu (0)=0$, the theorem indicates at
least one more zero at some finite value of $k_\mu$. The Green's function,
$(C_\mu (k)\sigma_\mu )^{-1}$, therefore has additional poles corresponding to
additional massless fermions. This is the origin of the problem [8].

To eliminate the unwanted states one might consider the modified Green's
function
\begin{equation}
G(k)^{-1}=C_\mu (k)\sigma_\mu +B(k)
\end{equation}
where the scalar function, $B(k)$, is chosen to vanish at $k=0$ but nowhere
else on the torus. For example, if $B(k)\sim rk^2$ then (1.3) would approximate
(1.1) near $k=0$ but there would be no other light fermions. Unfortunately,
this is not an acceptable solution. It leads indirectly to the violation of
Lorentz invariance in the low energy sector of the theory. Loop effects in
theories with interactions give contributions involving integrals over the
torus and these are, of course, quite generally not Lorentz invariant. However,
Lorentz invariance (or $SO(4)$ in the Euclidean version) is recovered,
approximately, in the low energy sector if the lattice theory possesses a
sufficiently strong discrete symmetry. One of the 4--dimensional hypercubic
crystal structures is generally adequate for this purpose. To recover chirality
in the low energy sector it is further necessary that the crystal group itself
admits chiral spinor representations (an example of this is the $F_4$ weight
lattice whose Weyl group has such 2--component spinor representations [9]). The
scalar term, $B(k)$, in (1.3) is not allowed by this kind of crystal symmetry.
With 2--component fermions there is nothing else to be done, the problem is
insoluble.

The approach of Narayanan and Neuberger uses 4--component fermions. Each Weyl
spinor is paired with an auxiliary spinor of opposite chirality. Instead of
working directly in 4--dimensional Euclidean spacetime they consider, following
Kaplan, an auxiliary quantum mechanics problem in $4+1$--dimensions. They
define
two Hamiltonians,
\begin{equation}
H_\pm =\int d^4x\ \psi (x)^\dagger\ \gamma_5(\gamma_\mu\
\partial_\mu\pm\vert\Lambda\vert )\psi (x)
\end{equation}
where $\gamma_\mu$ and $\gamma_5$ are hermitian Dirac matrices and $\Lambda$ is
an auxiliary parameter. The Schr\"odinger picture fields (from the
$4+1$--dimensional point of view) satisfy appropriate anticommutation rules,
$$
\{\psi (x), \psi^\dagger (x')\}=\delta_4 (x-x')
$$
etc. The eigenvalue spectra of $H_+$ and $H_-$ are identical but their
eigenstates are not. In particular, there are two distinct Dirac vacuum states,
$\vert +\rangle$ and $\vert -\rangle$, whose overlap, $\langle +\vert
-\rangle$, becomes an interesting quantity. Specifically, if the fermions are
coupled to an external gauge potential in the usual way,
$$
\partial_\mu\ \psi (x)\to (\partial_\mu -i\ A_\mu (x))\ \psi (x)
$$
then the overlap becomes a functional of $A$,
\begin{equation}
\langle A+\vert A-\rangle\ =\ \langle +\vert -\rangle\ e^{-\Gamma (A)}
\end{equation}
This functional will be interpreted, in the limit $\vert\Lambda\vert\to\infty$,
as the vacuum amplitude for a chiral fermion. Symbolically,
\begin{equation}
\ell n\ \det\left[{1+\gamma_5\over 2}\ (\partial\!\!\!/ -i\ A\!\!\!/ )\right]
\equiv -\lim_{\vert\Lambda\vert\to\infty}\ \Gamma (A)
\end{equation}

Is this a sensible interpretation? The answer is not yet clear. At present all
we can do is try to compute $\Gamma (A)$ and verify that it has the features to
be expected of a chiral vacuum amplitude, including the anomaly.

Some plausibility is found, we believe, by defining an appropriate free fermion
Green's function,
$$
G_\Lambda (x-x')={1\over\vert\Lambda\vert}\ {\langle +\vert\psi (x)\ \bar\psi
(x')\vert -\rangle\over\langle +\vert -\rangle}
$$
where $\bar\psi =\psi^\dagger\gamma_5$. In the limit this gives [5]
\begin{equation}
\lim_{\vert\Lambda\vert\to\infty}\ \widetilde G_\Lambda (k) = {1+\gamma_5\over
2}\ {1\over ik\!\!\!/}
\end{equation}
which is encouraging. It suggests that the positive chirality components of the
rescaled field $\vert\Lambda\vert^{-1/2}\ \psi (x)$ generate massless fermions
while the negative chirality components do not. Further plausibility is added
by perturbative computation of $\Gamma$, assuming $A_\mu$ to be weak. In
Sec.5 we examine the terms of second order and show that the expected
structure emerges. In Sec.4 we compute chiral anomalies in the 2-- and
4--dimensional cases.

Most of our computations are of course expressed in the lattice regularized
version of the theory since that is the real motivation for this work. The
1--body Hamiltonians,\break
$\widetilde H_\pm =\gamma_5(ik\!\!\!/ \pm \vert\Lambda\vert )$, are replaced by
\begin{equation}
\widetilde H_\pm (k) =\gamma_5\left[ i\ \gamma^\mu \ C_\mu
(k)+(B(k)\pm\vert\Lambda\vert )T_c\right]
\end{equation}
where $T_c$ is a diagonal matrix with $N_L$ eigenvalues $-1$, and $N_R$
eigenvalues $+1$. It commutes with the Dirac matrices and serves to label the
chiral flavours. The Hamiltonians (1.8) are invariant with respect to the
global group $U(N_L)\times U(N_R)$. (The breaking of this symmetry by a mass
term is considered in Sec.2.)

It should be emphasized at this point that $\Lambda$ has nothing to do with
ultraviolet regularization. The lattice takes care of that problem. Scales are
generally in the relation
$$
k\ll\Lambda\ll{1\over a}
$$
where $k$ is a typical external momentum and $a$ is the lattice spacing. While
$\Lambda$ goes to infinity relative to $k$, it goes to zero relative to the
lattice scale $1/a$. We shall for the most part use a natural system of units
with $a=1$ in this paper. In these units $\Lambda$ goes to zero but $k/\Lambda$
is understood always to be small. In particular, the chiral anomaly, which
shows up as a relative phase in the gauge transformation of the ground states
$\vert A+\rangle$ and $\vert A-\rangle$, will be computed as the discontinuity
at $\Lambda =0$ of the angle functional. Of course, it is also possible to
obtain the anomaly by setting $a=0$, ignoring the ultraviolet divergences,
keeping $k$ finite and letting $\Lambda$ go infinity. This procedure while
not very satisfying from a mathematical point of view, is much simpler to
apply. It will be used for the evaluation of the bilinear terms in Sec.5 and
the 2--dimensional gravitational anomalies in Sec.6.

Throughout this paper the lattice, i.e. its geometry and for the most part its
dimensionality will be arbitrary. Our Hamiltonians (1.8) are restricted only by
the crystal symmetry of the lattice. In particular no further assumptions are
made about the nature of couplings, i.e. nearest neighbour next nearest, etc.

In this paper we have exclusively concentrated on a particular approach to the
problem of chiral fermions on the lattice. We should, however, emphasize that
there are other interesting studies of this difficult problem in the literature
[10].

The plan of this paper is as follows. Sec.2 treats the kinematics of free
fermions, Green functions and chiral symmetry breaking on the lattice. Sec.3
introduces the coupling of lattice fermions to slowly varying weak external
gauge fields and some perturbative formulae are derived. Sec.4 is devoted to
the
chiral anomalies. The cases of 2 and 4-dimensional lattices are discussed in
detail. In this section the criteria for the cancellation of anomalies on
arbitrary 4-dimensional lattices are derived. In Sec.5 we calculate the
bilinear
part of the overlap amplitude in a continuum approximation and show that it
coincides with the second order terms of the effective action of
$D$--dimensional Weyl fermions coupled to an external gauge field. In Sec.6 we
apply the continuum version of the overlap formalism to the computation of the
Lorentz, diffeomorphism and Maxwell anomalies in the coupling of the Weyl
fermions to 2--dimensional gravitational and electromagnetic fields. Some
technical details of analysis of the relation between the doublers  and chiral
anomalies have been relegated to an appendix.

\section{Free fermions}

As described above the idea is to represent chiral fermions in $D$--dimensional
Euclidean spacetime in terms of Dirac fermions in a $D+1$--dimensional
Minkowskian space. We shall deal mainly with arbitrary $D$ but, for
simplicity, some of the features will be illustrated in $D=2$ and $4$. To each
chiral spinor in $D$--dimensions is associated a Dirac spinor in
$D+1$--dimensions. The number of field components is doubled. Moreover, in the
$D+1$--dimensional ``parent'' theory the fields depend on the extra unphysical
coordinate and the number of degrees of freedom is thereby enlarged. To make a
precise association of $D+1$--dimensional fields with $D$--dimensional degrees
of freedom
 a regulator mass,
$\Lambda$, is introduced in the parent Hamiltonian and a prescription is found
for scaling the fields and taking the limit
$p/\Lambda\to 0$ that projects onto chiral states with 4--momentum $p_\mu$.

Although the required prescription could be formulated in continuum field
theory we shall deal with fermions on a lattice since this is the real
motivation behind this formalism. In the lattice formulation fermionic
variables are associated with the sites of an integer lattice in $D$
dimensions, $\psi = \psi (n), n\in Z\!\!\! Z^D$. The fields have $2^{D/2}$
components and they satisfy the usual anticommutation rules,
\setcounter{equation}{0}
\renewcommand{\theequation}{2.\arabic{equation}}
\begin{eqnarray}
\{\psi_\alpha (n), \psi_\beta (m)\} &=& 0\ ,\nonumber \\
\{\psi_\alpha (n), \psi_\beta (m)^*\} &=& \delta_{nm}\
\delta_{\alpha\beta}\nonumber \\
\{\psi_\alpha (n)^*, \psi_\beta (m)^*\} &=& 0
\end{eqnarray}
where $\psi_\alpha (n)^*$ denotes the hermitian conjugate of $\psi_\alpha (n)$.
The Fock vacuum is defined by
\begin{equation}
\psi_\alpha (n)\vert 0\rangle =0
\end{equation}

The free Hamiltonian is bilinear in $\psi$ and $\psi^*$,
\begin{equation}
H=\sum_{n,m}\ \psi (n)^\dagger\ H(n-m)\ \psi (m)
\end{equation}
where the 1--body Hamiltonian, $H(n-m)$, is a matrix in Dirac space.
Translation invariance is assumed. The Fourier transform of the 1--body
Hamiltonian is given by
\begin{equation}
\widetilde H(k)=\sum_n\ H(n)\ e^{-ikn}
\end{equation}
where $kn=k_\mu\ n^\mu$ and the sum ranges over all sites of the integer
lattice. Conversely,
\begin{eqnarray}
H(n) & = & {1\over\Omega}\ \sum_k\ \widetilde H(k)\ e^{ikn}\nonumber \\
&\to &\int_{BZ} \left({dk\over 2\pi}\right)^D\ \widetilde H(k)\ e^{ikn}
\end{eqnarray}
where $\Omega$ denotes the number of sites on the lattice and the latter
expression corresponds to the limit $\Omega\to\infty$. If $\Omega$ is finite
then the sum over $k_\mu$ contains $\Omega$ terms. When $\Omega$ goes to
infinity
the resulting integral ranges over a cell of volume $(2\pi )^D$. Since
$\widetilde H(k)$ is periodic it is defined, in effect, on a torus. The momenta
range from $-\pi$ to $\pi$ or, equivalently, over a suitably constructed
Brillouin zone.

In addition to translation invariance it is necessary in practice to assume a
crystal symmetry. Although it may not be the most general structure allowed by
the symmetry, we shall adopt the form
\begin{equation}
\widetilde H(k) = \gamma_5 \Bigl( i\gamma^\mu\ C_\mu (k) +B(k)+\Lambda \Bigr)
\end{equation}
where $\gamma_\mu$ and $\gamma_5$ are hermitian Dirac matrices.
\begin{eqnarray}
\{\gamma^\mu ,\gamma^\nu\} & = & 2g^{\mu\nu}\nonumber \\
\{\gamma^\mu, \gamma_5\} & = & 0\nonumber \\
\gamma^2_5 & = & 1
\end{eqnarray}
The metric tensor $g^{\mu\nu}$ is an invariant tensor in the sense that
$g_{\mu\nu}\ n^\mu\ m^\nu$ is invariant with respect to the point group of the
lattice. The coefficients, $C_\mu (k), B(k)$ and $\Lambda$ are real
\footnote{ For example, a simple cubic lattice with nearest
neighbour couplings would have $g_{\mu\nu}=\delta_{\mu\nu},$\break
$ C_\mu (k)=\sin
k_\mu$ and $B(k)=r\ \mathop{\Sigma}\limits_\mu (1-\cos k_\mu )$.}.

In the natural units employed here, the lattice spacing is equal to unity and
the momentum components are angular variables. In these units the physically
interesting momenta are concentrated around $k=0$. The crystal symmetry forces
the vector $C_\mu$ to vanish at the origin and it can be normalized such that
\setcounter{equation}{0}
\renewcommand{\theequation}{2.8\alph{equation}}
\begin{equation}
C_\mu (k) \simeq k_\mu +\dots
\end{equation}
near $k=0$. By separating out the constant $\Lambda$ we may assume
\begin{equation}
B(k)\simeq rk^2+\dots
\end{equation}
where $k^2 =g^{\mu\nu}\ k_\mu\ k_\nu$ and $r$ is a constant. The $SO(D)$
invariant structure (2.8) around $k=0$ is ensured by imposing a sufficiently
potent crystal symmetry.

The constant $\Lambda$ will play an auxiliary role in projecting out the zero
modes. It will be taken to be large compared to the typical momentum $k_\mu$,
but small relative to the ultraviolet cutoff,
\setcounter{equation}{8}
\renewcommand{\theequation}{2.\arabic{equation}}
\begin{equation}
k_\mu\ll\vert\Lambda\vert\ll 1
\end{equation}
The sign of $\Lambda$ is significant and in the following we shall often
indicate it explicitly, for example by writing
\begin{equation}
\widetilde H_\pm (k) =\gamma_5 \Bigl( i\gamma^\mu\ C_\mu (k) +B(k)\pm
\vert\Lambda\vert \Bigr)
\end{equation}

To discuss the Hilbert space of the $D+1$--dimensional theory it is useful to
define creation and annihilation operators for the free fermions and their
antiparticles by making a plane wave expansion in the Schr\"odinger picture,
\begin{equation}
\psi (n) ={1\over\Omega}\ \sum_{k,\sigma}\left( b_\pm (k,\sigma )\ u_\pm
(k,\sigma )+ d^\dagger_\pm (k,\sigma )\ v_\pm (k,\sigma )\right) e^{ikn}
\end{equation}
where $u_\pm$ and $v_\pm$ denote the positive and negative energy eigenspinors
of the two Hamiltonians (2.10),
$$
\widetilde H_\pm (k)\ u_\pm (k,\sigma )=\omega_\pm (k)\ u_\pm (k,\sigma ),\ \ \
\widetilde H_\pm (k)\ v_\pm (k,\sigma )=-\omega_\pm\ v_\pm (k,\sigma )
$$
where
\begin{equation}
\omega_\pm (k) =\sqrt{ C_\mu (k)\ C_\mu (k) +(B(k)\pm \vert\Lambda\vert )^2}
\end{equation}
Since $\widetilde H_+$ and $\widetilde H_-$ are hermitian matrices the two sets
$(u_+,v_+)$ and $(u_-,v_-)$ are orthonormal and complete. A convenient choice
is given by
\begin{eqnarray}
u_\pm (k,\sigma ) & = & {\omega_\pm +B\pm\vert\Lambda\vert -iC\!\!\!\!/\over
\sqrt{2\omega_\pm  (\omega_\pm +B\pm\vert\Lambda\vert )}}\ \chi (\sigma
)\nonumber \\
v_\pm (k,\sigma ) & = & {\omega_\pm -B\mp \vert\Lambda\vert +iC\!\!\!\!/\over
\sqrt{2\omega_\pm (\omega_\pm -B\mp\vert\Lambda\vert )}}\ \chi (\sigma )
\end{eqnarray}
where $\gamma_5\chi (\sigma )=\chi (\sigma ),\ \  \chi^+(\sigma )\chi (\sigma
')
=\delta_{\sigma\sigma '}$, and the spin lable, $\sigma $, takes the values
$1,2,\dots ,2^{{D\over 2}-1}$. It is straightforward to find the relation
between
the two sets of eigenspinors,
\begin{eqnarray}
u_-(k,\sigma ) & = & \cos\beta\ u_+(k,\sigma )-\sin\beta\ v_+(k,\sigma
)\nonumber \\
v_-(k,\sigma ) & = & \sin\beta\ u_+(k,\sigma )+\cos\beta\ v_+(k,\sigma )
\end{eqnarray}
where the angle $\beta (k)$ is given by
\begin{eqnarray}
\sin\beta (k) & = & \sqrt{{\omega_++B+\vert\Lambda\vert\over 2\omega_+}\
{\omega_--B+\vert\Lambda\vert\over 2\omega_-}} -
\sqrt{{\omega_+-B-\vert\Lambda\vert\over 2\omega_+}\
{\omega_-+B-\vert\Lambda\vert\over 2\omega_-}}\nonumber \\
\cos\beta (k) & = &
\sqrt{{\omega_+-B-\vert\Lambda\vert\over 2\omega_+}\
{\omega_--B+\vert\Lambda\vert\over 2\omega_-}}
+\sqrt{{\omega_++B+\vert\Lambda\vert\over 2\omega_+}\
{\omega_-+B-\vert\Lambda\vert\over 2\omega_-}}\nonumber \\
&&
\end{eqnarray}
This angle lies between $-{\pi\over 2}$ and ${\pi\over 2}$. It is independent
of $\sigma$.

The anticommutors of $b_+,d_+$ and their hermitian conjugates take the usual
form. The only non--vanishing ones are
$$
\{ b_+(k,\sigma ),b^\dagger_+(k',\sigma ')\} =\Omega\ \delta_{kk'}\
\delta_{\sigma\sigma '} =\{ d_+(k,\sigma ), d^\dagger_+(k',\sigma ')\}
$$
Likewise for $b_-,d_-$. The two sets are related by a Bogoliubov
transformation,
\begin{eqnarray}
b_-(k,\sigma ) & = & \cos\beta\ b_+(k,\sigma )-\sin\beta\
d^\dagger_+(k,\sigma )\nonumber \\
d^\dagger_-(k,\sigma ) & = & \sin\beta\ b_+(k,\sigma )+\cos\beta\
d^\dagger_+(k,\sigma )
\end{eqnarray}
so that, for example,
\begin{eqnarray}
\{ d_+(k,\sigma ), d^\dagger_-(k',\sigma ')\} & = & \cos\beta\ \delta_{kk'}\
\delta_{\sigma\sigma '}\nonumber \\
\{ d_+ (k,\sigma ),b_-(k',\sigma ')\} & = & -\sin\beta\ \delta_{kk'}\
\delta_{\sigma\sigma '}
\end{eqnarray}

The Fock vacuum, $\vert 0\rangle$, is annihilated by $b_\pm$ and
$d^\dagger_\pm$. By filling the negative energy states one defines two Dirac
vacua,
\begin{equation}
\vert\pm\rangle =\mathop{\Pi}\limits_{k,\sigma}\ \Omega^{-1/2}\ d_\pm (k,\sigma
)\vert 0\rangle
\end{equation}
These vacuum states are annihilated by $d_\pm (k,\sigma )$, respectively. They
are normalized and their overlap is given by
\begin{equation}
\langle +\vert -\rangle = \mathop{\Pi}\limits_k\ (\cos\beta (k))^\nu
\end{equation}
where $\nu =2^{{D\over 2}-1}$ is the number of spin states. Zeroes of the
overlap, due to the vanishing of one or more of the factors $\cos\beta (k)$,
will be interpreted as a signal that fermion zero--modes are present and these
will be associated with physical states of $D$--dimensional chiral theory. To
make this concrete we construct a 2--point Green's function.

The fermion Green's function is defined by the ratio
\begin{equation}
G(n-m)={\langle +\vert\psi (n)\ \psi (m)^\dagger \vert -\rangle\over\langle
+\vert -\rangle }
\end{equation}
It can be evaluated by substituting the plane wave expansions (2.11), using the
commutation rules and the vacuum conditions, $d_-\vert -\rangle = 0 =\langle
+\vert d^\dagger_+$,
\begin{eqnarray*}
G(n-m) & = & {1\over\Omega^2}\ \sum\ u_+(k,\sigma )\ {\langle +\vert
b_+(k,\sigma )\ b^\dagger_-(k'\sigma ')\vert -\rangle\over\langle +\vert
-\rangle}\ u^\dagger_- (k'\sigma ')\ e^{ikn-ik'm}\\
& = & {1\over\Omega}\ \sum_k\ \widetilde G(k)\ e^{ik(n-m)}
\end{eqnarray*}
where, to obtain $\widetilde G(k)$, one needs the matrix element
\begin{eqnarray*}
{\langle +\vert b_+(k,\sigma )\ b^\dagger_-(k',\sigma )\vert -\rangle\over
\langle +\vert -\rangle} & = & \Omega\ \delta_{kk'}\ \delta_{\sigma\sigma '}\
\cos\beta -
{\langle +\vert b^\dagger_-(k',\sigma ')\ b_+(k,\sigma )\vert
-\rangle\over\langle +\vert -\rangle}\\
& = & \Omega\ \delta_{kk'}\ \delta_{\sigma\sigma '}\
\cos\beta
+\sin\beta\ \sin\beta '\ {\langle +\vert d_+(k',\sigma ')\
d^\dagger_-(k,\sigma )\vert -\rangle\over\langle +\vert -\rangle }\\
& = & \Omega\ \delta_{kk'}\ \delta_{\sigma\sigma '}\ \left(\cos\beta +
{\sin^2\beta\over\cos\beta}\right)
\end{eqnarray*}
Hence,
\begin{equation}
\widetilde G(k) ={1\over \cos\beta (k)}\ \sum_\sigma\ u_+(k,\sigma )\
u^\dagger_-(k,\sigma )
\end{equation}
The polarization sum could be expressed in terms of $C_\mu ,B$ and
$\vert\Lambda\vert$ using the explicit formulae (2.13) but the result is not
very informative. Instead we shall explore the vicinity of $k=0$. From (2.8)
and (2.12) we have
$$
\omega_\pm\simeq\vert\Lambda\vert +{k^2\over 2\vert\Lambda\vert} \pm rk^2
+\dots
$$
for $k\ll\vert\Lambda\vert\ll 1$. It follows from (2.13) and (2.15) that
\begin{eqnarray*}
\cos\beta &\simeq& \sqrt{{k^2\over\Lambda^2}}\ (1+\dots )\\
u_+(k,\sigma ) &\simeq&\left( 1-{ik\!\!\!/\over 2\vert\Lambda\vert}
+\dots\right)\chi (\sigma )\\
u_-(k,\sigma )&\simeq& \left( -{ik\!\!\!/\over\sqrt{k^2}}+{\sqrt{k^2}\over
2\vert\Lambda\vert} +\dots\right) \chi(\sigma )
\end{eqnarray*}
which can be substituted into (2.21) to give
\begin{equation}
\widetilde G(k)\simeq\vert\Lambda\vert\ {1+\gamma_5\over 2}\ {ik\!\!\!/\over
k^2} +\dots
\end{equation}
The vanishing of $\cos\beta$ at $k=0$ is thereby seen to mimic the contribution
of a massless fermion with chirality, $\gamma_5=+1$. To normalize this
contribution it would be necessary to scale the fields, $\psi
(n)\to\vert\Lambda\vert^{-1/2}\ \psi (n)$.

One may ask whether there are any other points on the momentum torus where
$\cos\beta$ vanishes. This is the famous doubling problem for chiral fermions
on a lattice. Since, for real $k_\mu$ the square roots in the expression (2.15)
are non--negative, they must both vanish if $\cos\beta$ is to vanish. This
implies
$$
\omega_+=\Big\vert B+\vert\Lambda\vert\Big\vert\quad {\rm and}\quad
\omega_-=\Big\vert B-\vert\Lambda\vert\Big\vert
$$
or, according to (2.12), $C_\mu =0$. However, this is not sufficient. It is
also
necessary that $B+\vert\Lambda\vert$ and $B-\vert\Lambda\vert$ should have
opposite signs. Necessary and sufficient conditions for the vanishing of
$\cos\beta$ are
\begin{equation}
C_\mu (k) =0\quad {\rm and}\quad B(k)^2<\Lambda^2
\end{equation}
These conditions are satisfied at $k=0$. To avoid the unwelcome doubling of
massless fermions they must be violated for $k\not= 0$. According to the
Poincar\'e--Hopf theorem there are other points on the momentum torus where
$C_\mu =0$. However, if $B^2>\Lambda^2$ at these points then $\cos\beta$ will
not vanish. Hence, to exclude the unwanted states one has only to choose a
function $B(k)$ that vanishes at $k=0$ but nowhere else, and take
$\vert\Lambda\vert$ sufficiently small. If this is done then the infrared
structure of the theory will be strictly confined to the neighborhood of $k=0$.

Up to this point we have been considering a single Dirac fermion, $\psi_\alpha
(n)$, on the lattice. Generalization is straightforward. Replace the
Hamiltonians (2.10) by the matrices
\begin{equation}
\widetilde H_\pm (k)^{ij}=\gamma_5\left\{ i\gamma^\mu C_\mu (k)\
\delta^{ij}+\left( B(k)\pm\vert\Lambda\vert\right)\ T^{ij}_c\right\}
\end{equation}
where the (flavour) indices $i,j$ run from 1 to $N_L+N_R$ and
$T_c$ is a diagonal matrix with $N_L$ entries $-1$ and $N_R$ entries $+1$.
Define the Dirac vacua and the Green's function exactly as before. Most of the
above formulae can be adapted by making the replacement $B\pm
\vert\Lambda\vert\to (B\pm\vert\Lambda\vert )T_c$ and it is understood that the
spinors $\chi (\sigma )$ now carry a flavour index and they are required to
satisfy
\begin{equation}
\gamma_5\ T_c\ \chi (\sigma )=\chi (\sigma )
\end{equation}
Equations (2.13) are replaced by
\begin{eqnarray}
u_\pm (k,\sigma ) & = & {\omega_\pm +B\pm\vert\Lambda\vert
-iC\!\!\!\!/\ T_c\over
\sqrt{2\omega_\pm  (\omega_\pm +B\pm\vert\Lambda\vert )}}\ \chi (\sigma
)\nonumber \\
&&\nonumber \\
v_\pm (k,\sigma ) & = & {\omega_\pm -B\mp \vert\Lambda\vert
+iC\!\!\!\!/\ T_c\over
\sqrt{2\omega_\pm (\omega_\pm -B\mp\vert\Lambda\vert )}}\ \chi (\sigma )
\end{eqnarray}
and, near $k=0$, (2.22) is replaced by
\begin{equation}
\widetilde G(k)\simeq\vert\Lambda\vert\ {1+\gamma_5 T_c\over 2}\
{ik\!\!\!/\over k^2}\ T_c
\end{equation}
The pole at $k^2=0$ now corresponds to a massless fermion with chirality
$\gamma_5\ T_c=1$. The peculiar factor, $T_c$, on the right of (2.27) can be
removed if we define the adjoint field
\begin{equation}
\bar\psi (n) =\psi (n)^\dagger\ \gamma_5
\end{equation}
This is appropriate because $\gamma_5$ plays a role in the $4+1$--dimensional
theory analogous to $\gamma_0$ in $3+1$--dimensional theory. With this
definition the result (2.27) would be replaced by the form appropriate to
chiral
fermions in Euclidean spacetime,
\begin{equation}
\widetilde G(k)\gamma_5\simeq\vert\Lambda\vert\ {1+\gamma_5T_c\over 2}\ {1\over
ik\!\!\!/}
\end{equation}
near $k=0$. The left (right) flavours are distinguished by $T_c=-1 (+1)$.

Up to this point we have described a system that represents massless chiral
fermions in $D$--dimensional spacetime with a global chiral symmetry
$U(N_L)\times U(N_R)$. To complete this section we consider the breaking of
chiral symmetry by mass terms. We can suppose that the symmetry breaking is
spontaneous, i.e. represent the mass by the expectation value of a Higgs field,
although we shall not discuss the Higgs dynamics. Let the Hamiltonian (2.3) be
modified by the addition of a Yukawa term.
\setcounter{equation}{29}
\renewcommand{\theequation}{2.\arabic{equation}}
\begin{equation}
\sum_n\ g\ \psi (n)^\dagger\ \gamma_5\ \phi (n)\ \psi (n)
\end{equation}
where $\phi (n)$ belongs to the representation $(N_L,\bar N_R)\oplus (\bar N_L,
N_R)$, i.e. an $N_L+N_R$ hermitian matrix with zeroes in the $N_L\times N_L$
and $N_R\times N_R$ blocks. Suppose it acquires a constant vacuum expectation
value
\begin{equation}
g\langle \phi (n)\rangle =m
\end{equation}
The 1--body Hamiltonian (2.24) is thereby modified to read
\begin{equation}
\widetilde H(k) =\gamma_5 (ik\!\!\!/ +m)\pm\vert\Lambda\vert\gamma_5\ T_c
+\dots
\end{equation}
near $k=0$. We take $m\ll\vert\Lambda\vert$ and assume that $B(k)$ has been
chosen to eliminate any other light fermions. Our aim is to compute the Green's
function near $k=0$ by expanding in powers of $k/\Lambda$ and $m/\Lambda$,
treating $k$ and $m$ as comparable. For this perturbative calculation it is
convenient to introduce a new set of eigenspinors in place of (2.13).

Write (2.32) as the sum of zero and first order terms
\begin{eqnarray}
\widetilde H_\pm (k) & = & H_{0\pm} + V(k)\nonumber \\
H_{0\pm} & = & \pm\vert\Lambda\vert\gamma_5\ T_c \nonumber \\
V(k) & = & \gamma_5(ik\!\!\!/ +m)
\end{eqnarray}
Note that $H_{0\pm}$ anticommutes with $V(k)$ because $\gamma_5$ anticommutes
with $k\!\!\!/$ and $T_c$ anticommutes with $m$. Define the zeroth order
eigenspinors $\chi_\pm (\sigma )$ such that
\begin{equation}
\gamma_5\ T_c\ \chi_\pm (\sigma ) =\pm\chi_\pm (\sigma )
\end{equation}
where $\sigma$ represents spin and flavour. It is easy to verify that, to first
order in $V/\vert\Lambda\vert$,
\begin{eqnarray}
u_\pm (k,\sigma ) & = & \left( 1+{1\over 2\vert\Lambda\vert}\
V+\dots\right)\chi_\pm (\sigma )\nonumber\\
v_\pm (k,\sigma ) & = & \left( 1-{1\over 2\vert\Lambda\vert}\ V+\dots \right)
\chi_\mp (\sigma )
\end{eqnarray}
where $u_\pm$ correspond to the eigenvalue $\vert\Lambda\vert$ of $\widetilde
H_\pm$ and $v_\pm$ correspond to $-\vert\Lambda\vert$. The various overlaps are
\begin{eqnarray}
u^\dagger_+(k,\sigma )\ u_-(k,\sigma ') & = & {1\over\vert\Lambda\vert}\
\chi^\dagger_+(\sigma )\ V(k)\chi_-(\sigma ')+\dots\nonumber \\
u^\dagger_+(k,\sigma )\ v_-(k,\sigma ') & = & \delta_{\sigma\sigma
'}+\dots\nonumber \\
v^\dagger_+(k,\sigma )\ u_-(k,\sigma ') & = & \delta_{\sigma\sigma
'}+\dots\nonumber \\
v^\dagger_+(k,\sigma )\ v_-(k,\sigma ') & = & -{1\over\vert\Lambda\vert}\
\chi^\dagger_-(\sigma )\ V(k)\ \chi_+(\sigma ')+\dots
\end{eqnarray}
where the dots represent terms of second order.

To compute the Green's function one needs the matrix element
\begin{eqnarray}
&&{\langle +\vert b_+(k,\sigma )\ b^\dagger_-(k',\sigma ')\vert -\rangle\over
\langle +\vert -\rangle }  =  \Omega\ \delta_{kk'}\ u^\dagger_+(k,\sigma )\
u_-(k,\sigma ')-{\langle +\vert b^\dagger_-(k',\sigma ')\ b_+(k,\sigma )\vert
-\rangle\over\langle +\vert -\rangle}\nonumber \\
&&\hspace{4.6cm} =  \Omega\ \delta_{kk'}\ u^\dagger_+(k,\sigma )\ u_-(k,\sigma
')-\nonumber\\ &&-\sum_{\sigma_1,\sigma_1'}\ u^\dagger_+(k,\sigma )\
v_-(k,\sigma_1)\ {\langle +\vert d_+(k',\sigma_1')d^\dagger_-(k,\sigma_1)\vert
-\rangle\over
\langle +\vert -\rangle}\ v^\dagger_+(k',\sigma_1')\ u_-(k',\sigma )
\end{eqnarray}
To evaluate the latter term use the commutation rules,
\begin{eqnarray*}
\{ d^\dagger_\pm (k,\sigma ), d_\pm (k',\sigma ')\} &=&\Omega\ \delta_{kk'}\
\delta_{\sigma\sigma '}\\
\{ d^\dagger_+ (k,\sigma ), d_- (k',\sigma ')\} &=&\Omega\ \delta_{kk'}\
K_{\sigma\sigma '}
\end{eqnarray*}
where
$$
K_{\sigma\sigma '}=v^\dagger_+(k,\sigma )\ v_-(k,\sigma ')
=-{1\over\vert\Lambda\vert}\ \chi^\dagger_-(\sigma )\ V(k)\ \chi_+(\sigma
')+\dots
$$
One obtains a ratio of determinants that expresses the inverse of $K$,
\begin{eqnarray}
{\langle +\vert d_+(k',\sigma_1')\ d^\dagger_-(k,\sigma_1)\vert
-\rangle\over\langle +\vert -\rangle} & = & \Omega\ \delta_{kk'}\
K^{-1}_{\sigma_1\sigma_1'}\nonumber \\
& = & -\Omega\ \delta_{kk'}\ \vert\Lambda\vert\ \chi^\dagger_+(\sigma_1)\
V(k)^{-1}\ \chi_-(\sigma_1 ')+\dots
\end{eqnarray}
where the dots indicate terms of zeroth order. Using this result together with
the formulae (2.36) one finds the leading term in (2.37),
$$
{\langle +\vert b_+(k,\sigma )\ b^\dagger_-(k',\sigma ')\vert
-\rangle\over\langle +\vert -\rangle} =\Omega\ \delta_{kk'}\ \vert\Lambda\vert\
\chi^\dagger_+(\sigma )\ V(k)^{-1}\ \chi_-(\sigma ')+\dots
$$
The Green's function follows immediately [5]
\begin{eqnarray*}
\widetilde G(k) & = & {1\over\Omega}\ \sum_{\sigma ,\sigma '}\
 u_+(k,\sigma )\ {\langle +\vert b_+(k,\sigma )\ b^\dagger_-(k,\sigma ')\vert
-\rangle\over\langle +\vert -\rangle}\ u^\dagger_-(k,\sigma ')\\
 & = & \vert\Lambda\vert\ {1+\gamma_5\ T_c\over 2}\ (ik\!\!\!/ +m)^{-1}\
\gamma_5+\dots
\end{eqnarray*}
The chiral symmetry breaking mass matrix, anticommuting with $T_c$, connects
left and right components as it should.

\section{Yang--Mills coupling}

The Hamiltonians $H_\pm$ discussed in Sec.2, for $N_L+N_R$ Dirac fermions on
the lattice, are invariant with respect to global $U(N_L)\times U(N_R)$. This
can be extended to local transformations by introducing Yang--Mills fields. To
couple an external gauge field we adopt a lattice version of the standard
minimal prescription. Define
\setcounter{equation}{0}
\renewcommand{\theequation}{3.\arabic{equation}}
\begin{equation}
H_\pm (A) = \sum_{n,m}\ \psi (n)^\dagger\ H_\pm (n-m)\ U(n,m)\ \psi (m)
\end{equation}
where $H_\pm (n-m)$ is the free 1--body Hamiltonian defined as the Fourier
transform of
\begin{equation}
\widetilde H_\pm (k) =\gamma_5\Biggl( i\gamma^\mu C_\mu
(k)+\Bigl( B(k)\pm\vert\Lambda\vert \Bigr) T_c\Biggr)
\end{equation}
and $U(n,m)$ is a unitary matrix in flavour space that commutes with the
chirality matrix, $T_c$. The matrix $U(n,m)$ is a functional of the external
gauge field, $A_\mu (x)$ which we assume to be smooth. It is defined by the
path--ordered integral,
\begin{eqnarray}
U(n,m) & = & T\left(\exp\ i\int^n_m A\right)\nonumber \\
& = & T\left(\exp\ i \int^1_0 dt\ \dot\xi^\mu (t)\ A_\mu (\xi (t))\right)
\end{eqnarray}
\begin{equation}
\xi^\mu (t) =t\ n^\mu +(1-t)m^\mu
\end{equation}
The path is a straight line joining the lattice sites $n$ and $m$. In the
following we shall always assume that $A_\mu (x)$ is weak and slowly varying.
The aim is to compute perturbative corrections to the vacuum amplitude $\langle
+\vert -\rangle$.

Under local transformations,
\begin{equation}
A_\mu (x)\to A^\theta_\mu (x) =e^{i\theta (x)}\ (A_\mu (x)+i\partial_\mu )
e^{-i\theta (x)}
\end{equation}
where $\theta (x)$ belongs to the algebra of $U(N_L)\times U(N_R)$, the
matrices $U$ transform according to
\begin{equation}
U(n,m)\to U^\theta (n,m)=e^{i\theta (n)}\ U(n,m) \ e^{-i\theta (m)}
\end{equation}
It follows that the Hamiltonian (3.1) transforms according to
\begin{eqnarray}
H_\pm (A^\theta ) & = & \sum_{n,m}\ \psi (n)^\dagger\ H_\pm (n-m)\ e^{i\theta
(n)}\ U(n,m)\ e^{-i\theta (m)}\ \psi (m)\nonumber \\
& = & U_\theta\ H_\pm (A)\ U^{-1}_\theta
\end{eqnarray}
where $U_\theta$ is a unitary operator that acts on the fermions,
\begin{eqnarray}
U_\theta\ \psi (n)\ U^{-1}_\theta & = & e^{-i\theta (n)}\ \psi (n)\nonumber \\
U_\theta\ \psi (n)^\dagger\ U^{-1}_\theta & = & \psi (n)^\dagger\ e^{i\theta
(n)}
\end{eqnarray}
It can be expressed in terms of these fields,
\begin{equation}
U_\theta =\exp\left( i\ \sum_n\ \psi (n)^\dagger\ \theta (n)\ \psi (n)\right)
\end{equation}

The perturbed vacuum states $\vert A\pm\rangle$ are obtained by solving the
eigenvalue problem
\begin{equation}
H_\pm (A)\vert A\pm\rangle\ = \vert A\pm\rangle\ E_\pm (A)
\end{equation}
or, writing $H_\pm (A) =H_\pm (0)+V$ and $E_\pm (A) =E_\pm (0)+\Delta E_\pm$,
$$
(E_\pm (0)-H_\pm (0))\vert A\pm\rangle\ = (V-\Delta E_\pm )\vert A\pm\rangle
$$
This can be expressed as an integral equation,
\begin{equation}
\vert A\pm\rangle\ = \vert\pm\rangle\ \alpha_\pm (A)+G_\pm (V-\Delta E_\pm
)\vert A\pm\rangle
\end{equation}
where $\vert\pm\rangle $ denotes the Dirac vacua of Sec.2 and
\begin{equation}
G_\pm ={1-\vert\pm\rangle\langle\pm\vert\over E_\pm (0)-H_\pm (0)}
\end{equation}
The vacuum energy shifts $\Delta E_\pm$ are determined by the consistency
requirement,
\begin{equation}
0=\langle\pm\vert (V-\Delta E_\pm )\vert A\pm\rangle
\end{equation}
The Dirac vacua $\vert\pm\rangle$ are non--degenerate ground states of the free
Hamiltonians $H_\pm (0)$ and, in perturbation theory at least, the interacting
ground states $\vert A\pm\rangle$ must also be non--degenerate. They are
expressed formally as solutions of (3.11),
\begin{equation}
\vert A\pm\rangle =\alpha_\pm (A)\left[ 1-G_\pm (V-\Delta E_\pm
)\right]^{-1}\vert\pm\rangle
\end{equation}
where the numerical factors $\alpha_\pm =\langle\pm\vert A\pm\rangle$ are
determined up to a phase by the normalization condition $\langle A\pm\vert
A\pm\rangle =1$. To fix the phase we choose $\alpha_\pm$ to be real and
positive.

In order to see to what extent the amplitude $\langle A+\vert A-\rangle$
resembles the vacuum amplitude for chiral fermions coupled to a weak external
gauge potential we shall calculate it in a continuum approximation to second
order in Sec.5. This is already a fairly cumbersome exercise and, although one
could extend it to higher orders, one could not expect to learn much from it
(except for possible anomalies to be considered in Sec.4).

Various matrix elements of the perturbation $V$ are needed. In particular,
\begin{eqnarray}
\langle 1\bar 2+\vert V\vert +\rangle & = & \langle +\vert\ d_+(2)\ b_+(1)\
V\vert +\rangle\nonumber\\
& = & \sum_{n,m}\ \langle +\vert d_+(2)\ b_+(1)\ \psi (n)^\dagger\
H(n-m)(U(n,m)-1)\psi (m)\vert +\rangle\nonumber \\
& = & \sum_{n,m}\ e^{-ik_1n+ik_2m}\ u^\dagger_+(1)\ H(n-m)(U(n,m)-1)v_+(2)
\end{eqnarray}
where
\begin{eqnarray}
U(n,m) & = & 1+i\int^1_0 dt_1\ \dot\xi^\mu (t_1)\ A_\mu (\xi (t_1))+\nonumber
\\
&&+{i^2\over 2}\int^1_0dt_1\ dt_2\ \dot\xi^\mu (t_1)\ \dot\xi^\nu (t_2)\
T\left( A_\mu (\xi (t_1)\ A_\nu (\xi (t_2))\right)
+\dots
\end{eqnarray}
To simplify the calculations we consider the infinite volume limit where the
momenta are continuous variables and
$$
H(n-m)=\int_{BZ}\left({dk\over 2\pi}\right)^D\ \widetilde H(k)\ e^{ik(n-m)}
$$
The integral extends over a Brillouin zone of volume $(2\pi )^D$. The external
field is smooth and slowly varying so that
$$
A_\mu (x) =\int \left({dk\over 2\pi}\right)^D\ \widetilde A_\mu (k)\ e^{ikx}
$$
where the integrand is concentrated around $k=0$. With these assumptions the
first order contribution reduces to
\begin{eqnarray}
\langle 1\bar 2+\vert V^{(1)}\vert +\rangle & = & -\sum_n\int^1_0 dt\
u^\dagger_+(1)\ \widetilde H_+\left( (1-t)k_1+tk_2-2\pi
nt\right)^{,\mu}\cdot\nonumber \\
&&\qquad\cdot\widetilde A_\mu (k_1-k_2+2\pi n)\ v_+(2)
\end{eqnarray}
where the lattice sum is needed to ensure periodicity in $k_1$ and $k_2$, and
$\widetilde H(p)^{,\mu} =\partial\widetilde H(p)/\partial p_\mu$. Since
$\widetilde A$ is concentrated at the origin only one term from this sum will
contribute for given values of $k_1$ and $k_2$. Also, since $k_1-k_2+2\pi n$ is
small it is practical to expand $\widetilde H$ in powers of $t$ and integrate
this parameter. One may also express (3.17) in the form
\begin{eqnarray}
\langle 1\bar 2 +\vert\ {\delta V\over\delta\widetilde A^\alpha_\mu (p)}\
\vert +\rangle & = & -(2\pi )^D\ \delta_{2\pi} (k_2-k_1+p)\cdot\nonumber \\
&&\cdot\int^1_0dt\ u^\dagger_+(1)\ \widetilde H_+(k_1-tp)^{,\mu}\ T_\alpha\
v_+(2)
\end{eqnarray}
where the matrices $T_\alpha$ provide a basis for the algebra and
$\delta_{2\pi}$ denotes the periodic delta function,
$$
\delta_{2\pi} (k_2-k_1+p)=\sum_n\ \delta (k_2-k_1+p+2\pi n)
$$
In the same notation the second order contribution is given by
\begin{eqnarray}
&&\langle 1\bar 2+\vert\ {\delta^2V\over\delta\widetilde A^\alpha_\mu
(p_1)\ \delta\widetilde A^\beta_\nu (p_2)}\vert +\rangle =\nonumber \\
&&\quad = (2\pi )^D\ \delta_{2\pi}(k_2-k_1+p_1+p_2)\int^1_0dt_1\ dt_2\
u^\dagger_+(1)\ \widetilde H(k_1-t_1p_1-t_2p_2)^{,\mu\nu}\cdot\nonumber \\
&&\quad\qquad\cdot\left\{\theta (t_1-t_2)\ T_\alpha T_\beta +\theta
(t_2-t_1)T_\beta T_\alpha\right\} v_+(2)
\end{eqnarray}
Again the integrals over $t_1$ and $t_2$ can be evaluated by expanding in
powers of the small momenta $p_1$ and $p_2$.

Other 2--particle matrix elements of $V$ are obtained from the expressions
(3.18) and (3.19) by using the appropriate eigenspinors in place of $u_+(1)$
and $v_+(2)$. For example, the matrix element $\langle 1+\vert V\vert
2+\rangle$ involves the replacement $v_+(2)\to u_+(2)$, whereas $\langle\bar
2+\vert V\vert \bar 1+\rangle$ requires $u^\dagger_+(1)\to -v^\dagger_+(1)$,
etc.
In these particular examples, however, if $k_1=k_2$ there will be a vacuum
contribution,
\begin{eqnarray}
\langle +\vert {\delta V\over\delta\widetilde A^\alpha_\mu (p)}\vert +\rangle &
= & -(2\pi )^D\ \delta_{2\pi}(p)\int_{BZ}\left({dk\over 2\pi}\right)^D\
\sum_\sigma\ v^\dagger_+(k)\ \widetilde H(k)^{,\mu}\ T_\alpha\ v_+(k)\\
\langle +\vert {\delta^2 V\over\delta\widetilde A^\alpha_\mu (p_1)\
\delta\widetilde A^\beta_\nu (p_2)}\vert +\rangle & = &
(2\pi )^D\ \delta_{2\pi} (p_1+p_2)\int^1_0 dt_1\ dt_2\cdot\nonumber \\
&&\cdot\int_{BZ}\left({dk\over 2\pi}\right)^D\ \sum_\sigma\ v^\dagger_+(k)\
\widetilde H(k-t_1p_1-t_2p_2)^{,\mu\nu}\{\theta (t_1-t_2)T_\alpha T_\beta
+\nonumber \\
&&\hspace{5cm} +\theta (t_2-t_1)T_\beta T_\alpha\}\ v_+(k)
\end{eqnarray}
The first order term (3.20) vanishes if the crystal symmetry is large enough
and we shall assume this is the case.

\section{Chiral anomalies}

The computation of perturbative corrections to the ground states was considered
in Sec.3. Here we wish to examine the response of these states to a gauge
transformation.

Gauge transformations are implemented by the unitary operators, $U_\theta$,
defined by (3.9). Acting on the Hamiltonians (3.1) they give
\setcounter{equation}{0}
\renewcommand{\theequation}{4.\arabic{equation}}
\begin{equation}
U_\theta\ H_\pm (A)\ U^{-1}_\theta = H_\pm (A^\theta )
\end{equation}
where $A^\theta$ is given by (3.5). Since the perturbative ground state is not
degenerate we must have
\begin{equation}
U_\theta\vert A\pm\rangle\ = \vert A^\theta\pm\rangle \ e^{i\Phi_\pm (\theta
,A)}
\end{equation}
where $\Phi_\pm$ is real. The ground state energies, $E_\pm (A)$, must be
invariant. The functionals $\Phi_\pm$ satisfy a composition rule that reflects
the group property,
\begin{equation}
e^{i\theta_1}\ e^{i\theta_2}=e^{i\theta_{12}}
\end{equation}
Applying the operators $U_{\theta_1}$ and $U_{\theta_2}$ successively to the
states $\vert A\pm\rangle$, using (4.2), one finds
\begin{equation}
\Phi_\pm (\theta_{12},A)=\Phi_\pm (\theta_1,A^{\theta_2})+\Phi_\pm (\theta_2,A)
\end{equation}
which must hold identically in $\theta_1,\theta_2$ and $A$ when $\theta_{12}$
is expressed in terms of $\theta_1$ and $\theta_2$. For infinitesimal
$\theta_1$ and $\theta_2$ (4.3) gives
$$
\theta_{12} =\theta_1+\theta_2+{i\over 2}\ [\theta_1,\theta_2]+\dots
$$
or, using a basis of hermitian matrices to write $\theta =\theta^\alpha \
T_\alpha$,
\begin{equation}
\theta^\alpha_{12} =\theta^\alpha_1+\theta^\alpha_2+{1\over 2}\
\theta^\gamma_1\ \theta^\beta_2\ c_{\beta\gamma}\ ^\alpha +\dots
\end{equation}
where $c_{\beta\gamma}\ ^\alpha$ is a structure constant. This formula is
accurate up to second order. Substituting the expansion,
$$
\Phi (\theta ,A) =\int dx\ \theta^\alpha (x)\ \Phi_\alpha (x\vert A)+{1\over
2}\int dx\
 dx'\
 \theta^\alpha (x)\ \theta^\beta (x')\ \Phi_{\alpha\beta}(x,x'\vert A)+\dots
$$
into the composition rule (4.4) one finds, in second order,
\begin{equation}
{1\over 2}\ \delta (x-x')\ c_{\alpha\beta}\ ^\gamma\ \Phi_{\gamma\pm}(x\vert
A) + \Phi_{\alpha\beta\pm}(x,x'\vert A)=
 -\nabla_\mu '\left(\delta\Phi_{\alpha\pm}(x\vert A)/\delta
A^\beta_\alpha (x')\right)
\end{equation}
where $\nabla_\mu$ denotes the covariant derivative $(A^\theta_\mu =A_\mu
+\nabla_\mu\theta +\dots )$.

The ground state overlap $\langle A+\vert A-\rangle$ may not be gauge
invariant. According to (4.2) it satisfies
\begin{equation}
\langle A^\theta +\vert A^\theta -\rangle \ =\ \langle A+\vert A-\rangle\
e^{ig(\theta ,A)}
\end{equation}
where $g(\theta ,A)=\Phi_+(\theta ,A)-\Phi_-(\theta ,A)$ may not vanish. This
is how the chiral anomaly is expressed in the overlap formulation. From the
composition formulae (4.6) it follows, in particular, that the first order part
of $g(\theta ,A)$ must satisfy the equations
\begin{equation}
\nabla_\mu\ {\delta\ g_\beta (x'\vert A)\over\delta\ A^\alpha_\mu
(x)}-\nabla_\mu '\ {\delta\ g_\alpha (x\vert A)\over\delta\ A^\beta_\mu (x')}=
\delta (x-x')\ c_{\alpha\beta}\ ^\gamma\ g_\gamma (x\vert A)
\end{equation}
which will be recognized as the Wess--Zumino consistency conditions. The
functional $g_\alpha (x\vert A)$ should therefore be interpreted as the
so--called ``consistent'' anomaly [11].

To compute the angle $\Phi_+$ it is sufficient to consider one component of the
defining equation (4.2),
$$
e^{i\Phi_+(\theta , A)}={\langle +\vert U_\theta\vert A+\rangle\over\langle
+\vert A^\theta _+\rangle}
$$
For infinitesimal $\theta$ this gives, using (3.9),
\begin{eqnarray}
\Phi_+(\theta ,A) & = & \sum_n\ {\langle +\vert \psi (n)^\dagger\ \theta (n)\
\psi (n)\vert A+\rangle\over \langle +\vert A+\rangle }+\nonumber \\
&&\quad +{1\over i}\int dx\ \theta^\alpha (x)\ \nabla_\mu\ {\delta\over\delta
A^\alpha_\mu (x)}\ \ell n\ \langle +\vert A+\rangle
\end{eqnarray}
Since $\langle +\vert A+\rangle =\alpha_+(A)$ is defined to be real and
positive
the second term here is pure imaginary. It makes no contribution to the real
angle,
$\Phi_+$. Hence, the first order part of $\Phi_+$ is given by
\begin{eqnarray}
\Phi_+(\theta ,A) & = & {\rm Re}\ \sum_n\ {\langle +\vert \psi (n)^\dagger\
\theta (n)\ \psi (n)\vert A+\rangle\over\langle +\vert A+\rangle }\nonumber \\
& = & {\rm Re}\ \sum_n\ \langle +\vert\psi (n)^\dagger\ \theta (n)\ \psi (n)
\left( 1- G_+(V-\Delta E_+)\right)^{-1}\vert +\rangle
\end{eqnarray}
on substituting the perturbation series (3.14). A similar expression gives
$\Phi_-(\theta ,A)$.

There is no reason to expect the quantity $g=\Phi_+-\Phi_-$ to vanish in
general. The Hamiltonians $\widetilde H_\pm (k)$ depend on the auxiliary
parameter $\Lambda$ through the combinations, $B(k)\pm\vert\Lambda\vert$,
respectively. There is no symmetry operator that relates them. However, it must
be kept in mind that $\Lambda$ is not an ultraviolet regulator. In the natural
units we are using, it is a small number. The physically significant features
of
the system are to be looked for in the infra--red sector,
$$
\vert k\vert\ll\vert \Lambda\vert\ll 1
$$
Amplitudes should be expanded in powers of $k$ and the coefficients should then
be evaluated at $\Lambda =0$. Singularities here correspond to ultraviolet
divergences and must be compensated in the usual way by counterterms. The
chiral anomaly, if it is present, will appear as a discontinuity in the angle
$\Phi$ at $\Lambda =0$. To illustrate the mechanism we consider firstly the
2--dimensional case, where the anomaly should come in the second order part of
the overlap $\langle A+\vert A-\rangle$ or, equivalently, in the first order
part of $\Phi_+-\Phi_-$. The 4--dimensional case will be considered below.

The first order contributions to $\Phi_\pm$ are easy to evaluate using the
matrix elements (3.18),
\begin{eqnarray}
\Phi^{(1)}_+(\theta ,A) & = & {1\over 2}\ \sum_n\ \langle +\vert\psi
(n)^\dagger\ \theta (n)\ \psi (n)\ G_+\ V^{(1)}\vert +\rangle + c.c.\nonumber
\\
& = & -{1\over 2}\ \sum_n \int\left({dk_1\over 2\pi}\right)^2\left({dk_2\over
2\pi}\right)^2\langle +\vert\psi (n)^\dagger\ \theta (n)\ \psi (n)\vert 1,\bar
2 +\rangle\ \left(\omega_+(1)+\omega_+(2)\right)^{-1}\cdot\nonumber \\
&&\cdot\int\left({dp\over 2\pi}\right)^2\ \langle 1,\bar 2+\vert\ {\delta
V\over\delta\widetilde A^\alpha_\mu (p)}\ \vert +\rangle\ \widetilde
A^\alpha_\mu
(p) +c.c.\nonumber \\
& = & {1\over 2}\int\left({dk_1\over 2\pi}\right)^2\left({dk_2\over
2\pi}\right)^2\ v^\dagger_+(2)\ \tilde\theta (2-1)\ u_+(1)\left(\omega_+(1)
+\omega_+(2)\right)^{-1}\cdot\nonumber \\
&&\cdot\int\left({dp\over 2\pi}\right)^2 (2\pi )^2\
\delta_{2\pi}(k_2-k_1+p)\int^1_0dt\ u^\dagger_+(1)\ \widetilde
H(k_1-tp)^{,\mu}\ T_\alpha\ v_+(2)\ \widetilde A^\alpha_\mu (p) +c.c.\nonumber
\\
& = & \int\left({dp\over 2\pi}\right)^2\ \tilde\theta_\alpha (-p)\ F^\mu_+(p)\
\widetilde A^\alpha_\mu (p)
\end{eqnarray}
where $F^\mu_+(p) =F^\mu_+(-p)^*$ is given by the loop integral,
\begin{eqnarray}
F^\mu_+(p)&=&{1\over 2}\int^1_0dt\int_{BZ}\left({dk\over 2\pi}\right)^2\
\left(\omega_+\Bigl( k+{p\over 2}\Bigr)+\omega_+\Bigl( k-{p\over
2}\Bigr)\right)^{-1}\cdot\nonumber \\
&&\cdot {\rm tr}\left( U_+\Bigl( k+{p\over 2}\Bigr)\ \widetilde H\left(
k+\Bigl({1\over 2}-t\Bigr) p\right)^{,\mu}\ V_+\Bigl( k-{p\over 2}\Bigr)
+\right.\nonumber \\
&&\quad \left.+V_+\Bigl( k+{p\over 2}\Bigr)\ \widetilde H\left(
k+\Bigl({1\over 2}-t\Bigr) p\right)^{,\mu}\ U_+\Bigl( k-{p\over
2}\Bigr)\right)
\end{eqnarray}
The Hamiltonians,
$$\widetilde H_\pm (k)=\gamma_5\ i\gamma^\nu\ C_\nu (k) +\left(
B(k)\pm\vert\Lambda\vert\right)\gamma_5\ T_c
$$
commute with the generators $T_\alpha$. The matrices $U_\pm , V_\pm$ project
onto positive and negative energy eigenstates of $\widetilde H_\pm$,
\begin{eqnarray*}
U_\pm (k) & = & {\omega_\pm (k)+\widetilde H_\pm (k)\over 2\omega_\pm (k)}\\
& = & 1-V_\pm (k)
\end{eqnarray*}
where $\omega_\pm =\sqrt{C^2+(B\pm \vert\Lambda\vert )^2}$. Since $U_+V_+=0$ it
is clear that (4.12) vanishes at $p=0$. The first derivative reduces to the
simple form
\begin{equation}
F^{\mu ,\nu}_\pm (0) ={1\over 16}\int_{BZ}\left({dk\over 2\pi}\right)^2\
{1\over\omega^3_\pm}\ {\rm tr}\left([\widetilde H^{,\mu},\widetilde
H^{,\nu}]\widetilde H_\pm\right)
\end{equation}
The trace over $2\times 2$ Dirac matrices gives, for each flavour,
\begin{eqnarray}
{\rm tr}\left( [\widetilde H^{,\mu},\widetilde H^{,\nu}]\widetilde H_\pm\right)
& = & 4i\ \varepsilon^{\alpha\beta}\ C_\alpha\ ^{,\mu}\ C_\beta\ ^{,\nu}\
(B\pm\vert\Lambda\vert )T_c
-4i\ \varepsilon^{\alpha\beta}\ (C_\alpha\ ^{,\mu}\ B^{,\nu}-C_\alpha\
^{,\nu}\ B^{,\mu})C_\beta\ T_c\nonumber \\
& = & 4i\ \varepsilon^{\mu\nu}\ T_c\Bigl\{ (B\pm\vert\Lambda\vert )\ \det
C-\bigr.
\bigl.  C_\alpha\ ^{,\alpha}\ B^{,\beta}\ C_\beta +C_\alpha\
^{,\beta}\ B^{,\alpha}\ C_\beta\Bigr\}
\end{eqnarray}
A discontinuity at $\Lambda =0$, i.e. $F^{\mu ,\nu}_+\not= F^{\mu ,\nu}_-$, can
arise in (4.13) from infra--red singularities, points where $\omega_\pm (k)=0$
at $\Lambda =0$. One such point is the origin since $C_\mu (k)\sim k_\mu$ and
$B(k)\sim k^2$ near $k=0$. In general one must sum the contributions from all
points where $C_\mu =B=0$. The contribution of one such point is evaluated in
the Appendix,
\begin{equation}
F^{\mu ,\nu}_+(0)-F^{\mu ,\nu}_-(0) ={1\over 4\pi}\ i\ \varepsilon^{\mu\nu}\
T_c\ {\det C\over\vert\det C\vert}
\end{equation}

If $B(p)$ was set equal to zero everywhere then there would be a contribution
from every zero of $C_\mu (k)$, the total for each flavour would then contain
the factor
$$
\sum_{zeroes}\ {\det C\over\vert\det C\vert} =0
$$
The vanishing of this sum is a consequence of the toroidal topology of momentum
space (Poincar\'e--Hopf theorem). If $B(p)\equiv 0$ then there is no anomaly.
On the other hand if $B(p)$ vanishes only at $p=0$, there can be an anomaly,
viz.
\begin{equation}
\Phi_+(\theta ,A)-\Phi_-(\theta ,A)=-{1\over 8\pi}\ \int dx\
\varepsilon^{\mu\nu}\ {\rm tr} (T_c\ \theta (x)\ F_{\mu\nu}(x))
\end{equation}

We repeat this calculation for the 4--dimensional case. Now we assume that
$B(k)$ has been chosen to eliminate any unwanted states, its only zero being at
$k=0$. The anomaly should appear in the second order part of $\Phi_+-\Phi_-$. A
straightforward application of the formulae developed in Sec.3 gives
\begin{eqnarray}
\Phi^{(2)}_+(\theta ,A) & = & {\rm Re}\ \sum_n\ \langle +\vert\psi (n)^\dagger\
\theta (n)\ \psi (n)\left( G_+V^{(2)}+G_+V^{(1)}G_+V^{(1)}\right)\vert
+\rangle\nonumber \\ & = & {\rm Re}\ {1\over\Omega^2}\ \sum_{1,2}\
v^\dagger_+(2)\ \tilde\theta (2-1)\ u_+(1)\langle 1,\bar 2+\vert\left(
G_+V^{(2)}+G_+V^{(1)}G_+V^{(2)}\right)\vert +\rangle\nonumber \\ & = & {1\over
2}\ {\rm Re}\int\left({dp\over 2\pi}\right)^4\left({dq\over 2\pi}\right)^4\
\tilde A^\alpha_\mu (p)\ \tilde A^\beta_\nu (q)\
\tilde\theta^\gamma (-p-q)\ R^{\mu\nu}_{\alpha\beta\gamma}(p,q)
\end{eqnarray}
where, in the limit $\Omega\to\infty$, the kernel
$R^{\mu\nu}_{\alpha\beta\gamma}$ is given by the 1--loop integral,
\begin{eqnarray}
&&R^{\mu\nu}_{\alpha\beta\gamma}(p,q)=\int_{BZ}\left({dk\over
2\pi}\right)^4\
\Bigl(\omega_+(k)+\omega_+(k-p-q)\Bigr)^{-1}\int^1_0 dt\ dt'\cdot\nonumber \\
&&\quad\cdot\Biggl[ -{\rm tr}\Biggl( U_+(k)\ \widetilde H(k-tp-t'q)^{,\mu\nu}\
V_+(k-p-q)\{\theta (t-t')T_\alpha T_\beta +\theta (t'-t)T_\beta T_\alpha\}
T_\gamma\Biggr) +\Biggr.\nonumber \\
 &&\quad
+{2\over\omega_+(k-p-q)+\omega_+(k-p)}\ {\rm tr}\Biggl( U_+(k)\
\widetilde H(k-tp)^{,\mu}\ U_+(k-p)\ \widetilde H(k-p-t'q)^{,\nu}\cdot\Biggr.
\nonumber \\
&&\hspace{9cm} \cdot\Biggl. V_+(k-p-q)T_\alpha T_\beta
T_\gamma\Biggr)-\nonumber\\
 &&\Biggl. -{2\over\omega_+(k)+\omega_+(k-p)}\ {\rm tr}\Biggl( U_+(k)\
\widetilde H(k-tp)^{,\mu}\ V_+(k-p)\ \widetilde H(k-p-t'q)^{,\nu}\cdot\nonumber
\\
&&\hspace{9cm} \cdot V_+(k-p-q)T_\alpha T_\beta T_\gamma\Biggr)\Biggr]\nonumber
\\
&&
\end{eqnarray}
The analogous formulae for $\Phi_-$ are obtained by replacing
$\Lambda$ with
$-\Lambda$ wherever it occurs.

Instead of examining every term in the somewhat daunting expression (4.18) we
shall assume that the discontinuity at $\Lambda =0$ is confined to components
that contain the antisymmetric tensor, $\varepsilon^{\kappa\lambda\mu\nu}$.
These parts are relatively easy to pick out. Notice, firstly, that the trace
involving $\widetilde H^{,\mu\nu}$ makes no contribution since it has too few
$\gamma$ matrices. The second and third terms in (4.18) make similar
contributions involving the Dirac trace
\begin{eqnarray}
&&{1\over 4}\ {\rm tr}\left(\widetilde H(k)\ \widetilde
H(k-tp)^{,\mu}\
\widetilde H(k-p)\ \widetilde H(k-p-t'q)^{,\nu}\ \widetilde H(k-p-q)\right)
=\nonumber \\
 &&= 4\ \varepsilon^{\kappa\lambda\sigma\tau}\Biggl[ (B(k)+\Lambda
)\ C_\kappa (k-tp)^{,\mu}\ C_\lambda (k-p)\ C_\sigma (k-p-t'q)^{,\nu}\ C_\tau
(k-p-q)-\nonumber \\
&&\quad -C_\kappa (k)\ B(k-tp)^{,\mu}\ C_\lambda (k-p)\
C_\sigma (k-p-t'q)^{,\nu}\ C_\tau (k-p-q)+\nonumber \\
&&\quad +C_\kappa (k)\
C_\lambda (k-tp)^{,\mu}\ \Bigl(B(k-p)+\Lambda \Bigr)\ C_\sigma
(k-p-t'q)^{,\nu}\
C_\tau (k-p-q) -\nonumber \\ &&\quad - C_\kappa (k)\ C_\lambda (k-tp)^{,\mu}\
C_\sigma (k-p)\ B(k-p-t'q)^{,\nu}\ C_\tau (k-p-q)+\nonumber \\
&&\quad \Biggl.
+C_\kappa (k)\ C_\lambda (k-tp)^{,\mu}\ C_\sigma (k-p)\ C_\tau
(k-p-t'q)^{,\nu}\
\Bigl( B(k-p-q)+\Lambda \Bigr)\Biggr]
\end{eqnarray}
plus terms not containing $\varepsilon$. Now expand. Since
$\tilde A(p), \tilde A(q)$ are concentrated around the origin it is feasible to
treat $p$ and $q$ as small variables. Moreover, since we are looking for the
discontinuity at
$\Lambda =0$ which is casued by the infra--red singularity at $k=0$, we can
assume that the small $\Lambda$ behaviour is dominated by the neighbourhood of
$k=0$. Expanding (4.19) in powers of $p,q$ and $k$ we need to retain only the
terms of second order,
\begin{eqnarray*}
&&4\ \varepsilon^{\kappa\lambda\sigma\tau}\Bigl( \Lambda\
\delta^\mu_\kappa (k-p)_\lambda\ \delta^\nu_\sigma (k-p-q)_\tau +\Bigr.\\
&&\qquad + k_\kappa\ \delta^\mu_\lambda\ \Lambda\ \delta^\nu_\sigma\ \
(k-p-q)_\tau +\\
 &&\qquad \Bigl. + k_\kappa\ \delta^\mu_\lambda\ \ (k-p)_\sigma\
\delta^\nu_\tau\
\Lambda\Bigr)\\
&&=4\ \Lambda\ \varepsilon^{\mu\nu\lambda\tau}\ p_\lambda\ q_\tau
\end{eqnarray*}
The relevant part of (4.18) therefore reduces to
\begin{eqnarray}
R^{\mu\nu}_{\alpha\beta\gamma}(p,q) &\simeq &
-{1\over 2}\ \Lambda\ \varepsilon^{\mu\nu\lambda\tau}\ p_\lambda\ q_\tau\ {\rm
tr} (T_cT_\alpha T_\beta T_\gamma )\int\left({dk\over 2\pi}\right)^4\
{1\over\omega^5}\nonumber \\ & = & -{1\over 24\pi^2}\
{\Lambda\over\vert\Lambda\vert}\
\varepsilon^{\mu\nu\lambda\tau}\ p_\lambda\ q_\tau\ {\rm tr} (T_cT_\alpha
T_\beta T_\gamma )
\end{eqnarray} near $\Lambda =0$. The anomaly is given by
\begin{eqnarray}
\Phi_+-\Phi_- & = & -{1\over 24\pi^2}\int\left({dp\over
2\pi}\right)^4\left({dq\over 2\pi}\right)^4\ \tilde A^\alpha_\mu (p)\ \tilde
A^\beta_\nu (q)\ \tilde\theta^\gamma (-p-q)\cdot\nonumber \\ &&\qquad\quad\cdot
\varepsilon^{\mu\nu\lambda\tau}\ p_\lambda\ q_\tau\ {\rm tr} (T_cT_\alpha
T_\beta T_\gamma )\nonumber \\ & = & {1\over 24\pi^2}\int d^4x\
\varepsilon^{\mu\nu\lambda\tau}\ {\rm tr}
\left( T_c\ \partial_\lambda\ A_\mu (x)\ \partial_\tau\ A_\nu\ \theta
(x)\right)
\end{eqnarray} to second order. Presumably the full non--Abelian structure is
enforced by the consistency equations (4.8).

To construct anomaly--free models it is necessary to combine chiral multiplets
in a  suitably way. In effect this means choose a chirality matrix, $T_c$, that
commutes with $\theta (x)$ and satisfies
$$
\varepsilon_{\kappa\mu\lambda\nu}\ {\rm tr}(T_c\ \theta (x)\ \partial_\kappa\
A_\mu\ \partial_\lambda\ A_\nu )=0
$$
For example for one family of quarks and leptons in the standard $SU(2)_L\times
U(1)_Y$ model choose $T^{lep}_c= diag (1,1,-1)$ and $T^{quark}_c=diag
(1,1,-1,-1)$ for each colour.

\section{The bilinear part of $\langle A+\vert A-\rangle$}

In the foregoing section it was proven that the lattice effective action
$\Gamma(A)=-ln\langle A+\vert A-\rangle$ exhibits the correct anomalous
behaviour under local gauge transformations. In this section we would like to
show the plausibility of (1.5), in any number of dimensions, up to the second
order terms in $A$.

 For slowly varying weak background gauge
fields we shall approximate the lattice formulation of the last three sections
with a continuum theory described by the set of equations (1.4) to (1.6).
The starting point will be the continuum Hamiltonians
\setcounter{equation}{0}
\renewcommand{\theequation}{5.\arabic{equation}}
\begin{equation}
H_\pm = \int d^Dx\ \psi^\dagger (x)\ \gamma_5\Biggl(\gamma_\mu (\partial_\mu
-i\
A_\mu )\pm \vert\Lambda\vert\Biggr)\ \psi (x)
\end{equation}
where $A_\mu$ is the Lie algebra valued vector potential.

To evaluate $\Gamma (A) =-\ell n\ \langle A+\vert A-\rangle$ up to second order
terms in $A$ we use the perturbative solution (3.14). It is not hard to show
that for finite values of $\vert\Lambda\vert$ we have the following
non--vanishing contributions
\begin{eqnarray}
{\langle A+\vert A-\rangle\over\langle +\vert -\rangle} &=&
1-{1\over 2}\ \langle +\vert VG^2_+V\vert +\rangle -{1\over 2}\ \langle -\vert
VG^2_-V\vert -\rangle +\nonumber \\
&&+{\langle +\vert VG_+VG_+\vert -\rangle\over\langle +\vert -\rangle} +
{\langle +\vert VG_+G_-V\vert -\rangle\over\langle +\vert -\rangle} +
{\langle +\vert G_-VG_-V\vert -\rangle\over\langle +\vert -\rangle }
+\dots
\end{eqnarray}
where $G_\pm$ are defined by (3.12) and
\begin{equation}
V=\int d^Dx\ \psi^\dagger (x)\ i\ A\!\!\!/ (x)\ \gamma_5\ \psi (x)
\end{equation}
Each term appearing in (5.2) can be expressed in terms of matrix elements of
$V$ in which up to 4--particle states can contribute. For example
\setcounter{equation}{0}
\renewcommand{\theequation}{5.4\alph{equation}}
\begin{equation}
\langle +\vert VG^2_+V\vert +\rangle ={1\over\Omega^2}\ \sum_{1,2}\
{\langle +\vert V\vert 1\bar 2 +\rangle\langle 1\bar 2 +\vert V\vert
+\rangle\over (\omega_1+\omega_2)^2}
\end{equation}
and
\begin{eqnarray}
{\langle +\vert VG_+VG_+\vert -\rangle\over\langle +\vert -\rangle}
&=&{1\over\Omega^4}\
\sum_{1,2,3,4}\ {\langle +\vert V\vert 1\bar 2+\rangle\langle 1\bar 2+\vert
V\vert 3\bar 4 +\rangle\langle 3\bar 4 +\vert -\rangle\over
(\omega_1+\omega_2)(\omega_3+\omega_4)\langle +\vert -\rangle }\nonumber \\
&&+{1\over\Omega^6}\ \sum_{1\dots 6}\ {\langle +\vert V\vert 1\bar 2
+\rangle\langle 1\bar 2+\vert V\vert 34\bar 5\bar 6 +\rangle\langle 34\bar
5\bar 6 +\vert -\rangle\over
4(\omega_1+\omega_2)(\omega_3+\omega_4+\omega_5+\omega_6)\langle +\vert
-\rangle}
\end{eqnarray}
etc. The notation used here is identical to that of Sec.3. In particular the
labels $1,2,\dots ,$ stand for momentum, spin and flavour.

Expressing $V$ and the states in terms of Fock space operators one can
calculate all the necessary matrix elements. Here we give one of them as an
example
\setcounter{equation}{4}
\renewcommand{\theequation}{5.\arabic{equation}}
\begin{eqnarray}
\langle 1\bar 2 +\vert V\vert 3\bar 4 +\rangle & = & \Omega\ \delta_{24}\
u^\dagger_+(1)\ i\ \tilde A\!\!\!/ (k_1-k_3)\ \gamma_5\ u_+(3)\nonumber \\
&&-\Omega\ \delta_{12}\ v^\dagger_+(4)\ i\ \tilde A\!\!\!/ (k_4-k_2)\ \gamma_5\
v_+(2)
\end{eqnarray}
where
\setcounter{equation}{0}
\renewcommand{\theequation}{5.6\alph{equation}}
\begin{eqnarray}
u_\pm (k,\lambda ) & = & {\omega (k) +\gamma_5(i\ k\!\!\!/\pm \vert\Lambda\vert
)\over\sqrt{2\omega (k)\ \bigl(\omega (k)\pm\vert\Lambda\vert\bigr)}}\ \chi
(\lambda )\\
v_\pm (k,\lambda ) & = & {\omega (k)-\gamma_5(i\ k\!\!\!/ \pm\vert\Lambda\vert
)\over\sqrt{2\omega (k)\ \big(\omega (k)\mp\vert\Lambda\vert\bigr)}}\ \chi
(\lambda )
\end{eqnarray}
and $\omega (k)=\sqrt{k^2+\Lambda^2}$.\\
Using these equations we can put equations such as (5.4) into the
form
\setcounter{equation}{6}
\renewcommand{\theequation}{5.\arabic{equation}}
\begin{equation}
\langle +\vert VG^2_+V\vert +\rangle = {1\over\Omega^2}\ \sum_{k_1,k_2}\
{{\rm tr} \Bigl( i\ \tilde A\!\!\!/ (k_2-k_1)\gamma_5U_+(k_1)i\ \tilde A\!\!\!/
(k_1-k_2)\gamma_5\ V_+(k_2)\Bigr)\over \Bigl(\omega (k_1)+\omega (k_2)\Bigr)^2}
\end{equation}
where
\begin{equation}
U_\pm (k) ={\omega (k) +\gamma_5(i\ k\!\!\!/ \pm\vert\Lambda\vert )\over
2\omega (k)} =1-V_\pm (k)
\end{equation}
Every term in (5.2) can be rewritten in a form similar to (5.7). These
expressions of course in general suffer from ultraviolet divergences. It would
be more desirable to carry out these calculations on the lattice regularized
version of the overlap. We shall come back to this problem in the future but,
for now we give only the more brief and intuitive continuum version. For the
time being let us assume that the continuum theory is regularized in some way
and examine the $\vert\Lambda\vert\to\infty$ limit of expressions like (5.7).
It is not hard to see that in this limit (5.7) vanishes. In fact the only
non--vanishing contribution to (5.2), in the limit of
$\vert\Lambda\vert\to\infty$, originate from the last three terms of this
equation. It can be shown that as $\vert\Lambda\vert\to\infty$, the overlap
given
by (5.2) reduces to
\setcounter{equation}{0}
\renewcommand{\theequation}{5.9\alph{equation}}
\begin{equation}
{\langle A+\vert A-\rangle\over\langle +\vert -\rangle} =1-\left({1\over 8}
+{1\over 4}+{1\over 8}\right)\int {d^Dp\over (2\pi )^D}\ \tilde A_\mu\ ^\alpha
(p)\
\Pi_{\mu\nu}(p)\ \tilde A^\alpha_\nu (-p)
\end{equation}
where
\begin{equation}
\Pi_{\mu\nu} =\int {d^Dk\over (2\pi )^D}\ {\rm tr}\left(\gamma_\mu\
{1+\gamma_5\over 2}\ {k\!\!\!/ -p\!\!\!/\over (k-p)^2}\ \gamma_\nu\
{1+\gamma_5\over 2}\ {k\!\!\!/\over k^2}\right)
\end{equation}
The numerical coefficients in the second term on the right--hand side of (5.9a)
indicate the contribution of the respective terms in (5.2).

Equations (5.9a) and (5.9b) are in agreement with the bilinear part of the
effective action of a Weyl fermion coupled to Yang--Mills fields in a
$D$--dimensional Euclidean space. Namely if we start from
\begin{eqnarray*}
e^{-\Gamma (A)} & = & \int (d\bar\psi\ d\psi ) \ e^{-S}\\
S & = & \int d^Dx\ \bar\psi (x) \left(\partial\!\!\!/ -i\ A\!\!\!/\
{1+\gamma_5\over 2}\right)\ \psi (x)
\end{eqnarray*}
and calculate $\Gamma (A)$ to second order in $A$  we obtain $\Gamma (A)
=-\ell n\ {\langle A+\vert A-\rangle\over\langle +\vert -\rangle}$, with
$\langle A+\vert A-\rangle$ given by (5.9). This exercise indicates that as
$\vert\Lambda\vert\to\infty$, the unphysical states not only decouple in the
free theory as shown in [5] but they also do so in the quantum loops of
the interacting theory, leaving behind only the contribution of massless
physical
chiral fermions. Together with the anomaly calculations this gives further
support to the identification~(1.6).

Another indication of the validity of the overlap approach is contained in a
recent note of Narayanan and Neuberger [12] in which the chiral determinant on
a
2--dimensional torus in the presence of non--trivial background Polyakov loop
variables are examined.

\section{Gravitational anomalies}

In this section we show that the overlap formalism correctly reproduces the
gravitational anomalies. We shall examine only continuous two--dimensional
theories but we believe that the extension of our calculations to higher
dimensions should be in principle straightforward, although more cumbersome.
Also, being concerned with the continuum theory, as we are in this section, we
shall consider the limit $\vert\Lambda\vert\to\infty$ of the overlap.

The Hamiltonians which, in the limit of $\vert\Lambda\vert\to\infty$, produce
the appropriate effective action for the coupling of 2--dimensional chiral
fermions to gravity are given by
\setcounter{equation}{0}
\renewcommand{\theequation}{6.\arabic{equation}}
\begin{equation}
H_\pm =\int d^2x\ \psi^\dagger (x)\ \sigma^3\Bigl(\sigma^a\ e_a\ ^\mu (x)\
\nabla_\mu\pm\vert\Lambda\vert\Bigr)\ \psi (x)
\end{equation}
where $\sigma^3,\sigma^a,\ a=1,2$ are the Pauli spin matrices and $e_a\ ^\mu
(x)$ are the components of a zwei--bein. The covariant derivative $\nabla_\mu$
is defined by
\begin{equation}
\nabla_\mu\psi =\left(\partial_\mu -{i\over 2}\ \omega_\mu\ \sigma_3-i\ A_\mu
-{1\over 2}\ \partial_\mu\ \ell n\ e\right)\ \psi (x)
\end{equation}
where $e=\det\ e_\mu\ ^a(x)$ and $A_\mu$ is a $U(1)$--Maxwell field. The spin
connection $\omega_\mu$ is given by
\begin{equation}
\omega_\mu =-{1\over 2}\ e_\mu\ ^a\ {\varepsilon^{\alpha\beta}\over e}\
\partial_\alpha\ e_\beta\ ^a
\end{equation}
where $\varepsilon^{\alpha\beta}=-\varepsilon^{\beta\alpha} =
\varepsilon_{\alpha\beta}$ and $\varepsilon^{12}=1$.

The 2--component spinor field $\psi (x)$ can undergo three independent local
transformations, namely
\begin{description}
\item{i)}
local frame rotations:\hspace{2.8cm} $\psi'(x)=e^{{i\over 2}\sigma_3\phi (x)}\
\psi (x)$\hfill
$(6.4a)$

\item{ii)}
$U(1)$--gauge transformations:\hspace{1.3cm} $\psi ' (x) = e^{i\theta (x)}\
\psi
(x)$\hfill $(6.4b)$

\item{iii)}
general coordinate transformations: $\psi '(x')={\displaystyle\left(\det \
{\partial x^\mu\over\partial x^{'\nu}}\right)^{1/2}}\ \psi (x)$\hfill $(6.4c)$
\end{description}

\noindent The transformation rule (6.4c) indicates that $\psi (x)$ is a scalar
density of weight $1/2$ under the diffeomorphism group.

It is known that the effective action of charged chiral fermions in a
gravitational background responds anomalously to all three groups of the above
transformations [13]. Our intention in this section is to recover this
anomalous
behaviour within the overlap formalism. To this end we shall assume that the
external fields are weak and evaluate the angles $\Phi_\pm$ introduced in
Sec.4 up to first order in the external fields and the parameters of
transformations. Writing
\setcounter{equation}{4}
\renewcommand{\theequation}{6.\arabic{equation}}
\begin{equation}
e_a\ ^\mu (x)=\delta_a\ ^\mu+h_a\ ^\mu (x)
\end{equation}
we need to evaluate
\begin{equation}
\Phi_+(A, h;\varphi ,\theta ,\xi )={1\over \alpha_+}\ {\rm Re}\int d^2_x\
\langle +\vert\psi^+(x)\ \Sigma (x)\ \psi (x)\vert A,h+\rangle
\end{equation}
where
\setcounter{equation}{0}
\renewcommand{\theequation}{6.7\alph{equation}}
\begin{equation}
\Sigma (x) = {1\over 2}\ \sigma_3\ \varphi (x) +\theta (x)+\xi^\lambda (x)\
P_\lambda (x)
\end{equation}
where the operators $P_\lambda (x)$ are the generators of the diffeomorphisms
and are defined by
\begin{equation}
\psi^\dagger (x)\ P_\lambda (x)\ \psi (x) ={i\over 2}\ \left(\partial_\lambda\
\psi^\dagger (x)\ \psi (x) -\psi^\dagger (x)\ \partial_\lambda\ \psi (x)\right)
\end{equation}
$\Phi_-$ can be obtained from $\Phi_+$ by changing $\vert\Lambda\vert$ to
$-\vert\Lambda\vert$.

Using the same notation as in Sec.3, it is easy to see that
\setcounter{equation}{0}
\renewcommand{\theequation}{6.8\alph{equation}}
\begin{eqnarray}
A_L &\equiv&{\delta\over\delta\varphi (x)}\
(\Phi_+-\Phi_-)\Big\vert_{\vert\Lambda\vert\to\infty}\nonumber \\
& = & {\rm Re}\left[{1\over\Omega^2}\ \sum_{k_1k_2}\ {1\over\omega_1+\omega_2}
\langle +\vert\psi^+(x)\ {\sigma_3\over 2}\ \psi (x)\vert k_1\bar k_2+\rangle
\langle +k_1\bar k_2\vert V\vert +\rangle - (\vert\Lambda\vert\to
-\vert\Lambda\vert )\right]_{{\Omega\to\infty\atop\vert\Lambda\vert\to\infty}}
\nonumber\\
&&\\
&&\nonumber \\
A_{Maxw.} &\equiv&{\delta\over\delta\theta (x)}\
(\Phi_+-\Phi_-)\Big\vert_{\vert\Lambda\vert\to\infty}\nonumber \\
 & = & {\rm
Re}\left[{1\over\Omega^2}\ \sum_{k_1k_2}\ {1\over\omega_1+\omega_2}
\langle +\vert\psi^+(x)\  \psi (x)\vert k_1\bar k_2+\rangle
\langle +k_1\bar k_2\vert V\vert +\rangle - (\vert\Lambda\vert\to
-\vert\Lambda\vert )\right]_{{\Omega\to\infty\atop\vert\Lambda\vert\to\infty}}
\nonumber\\
&&\\
&&\nonumber \\
T_\mu &\equiv&{\delta\over\delta\xi^\mu (x)}\
(\Phi_+-\Phi_-)\Big\vert_{\vert\Lambda\vert\to\infty}\nonumber \\
& = & {\rm
Re}\Bigl[{1\over\Omega^2}\ \sum_{k_1k_2}\ {1\over\omega_1+\omega_2}
\langle +\vert{i\over 2}\left(\partial_\mu\ \psi^\dagger (x)\ \psi
(x)-\psi^\dagger (x)\ \partial_\mu\ \psi (x)\right)\vert k_1\bar k_2+\rangle
\langle+ k_1\bar k_2\vert V\vert +\rangle -\Bigr.\nonumber \\
&&\hspace{9cm} \Bigl. - (\vert\Lambda\vert\to
-\vert\Lambda\vert )\Bigr]_{{\Omega\to\infty\atop\vert\Lambda\vert\to\infty}}
\end{eqnarray}
where
\setcounter{equation}{8}
\renewcommand{\theequation}{6.\arabic{equation}}
\begin{equation}
V=\int d^2x\ \psi^\dagger (x)\ \sigma_3\left\{ \sigma^a\ h_a\ ^\mu (x)\
\partial_\mu +\sigma^\mu\left( -{i\over 2}\ \omega_\mu\ \sigma_3-i\ A_\mu
+{1\over 2}\ \partial_\mu\ h\right)\right\} \psi (x)
\end{equation}
and $h(x)=\delta^a_\mu\ h_a\ ^\mu (x)$.

Using the notation developed in previous sections we can easily evaluate the
matrix elements appearing in (6.8). For example
\begin{eqnarray*}
\langle k_1\bar k_2+\vert V\vert +\rangle
& = & u^\dagger_+(k_1)\
\sigma_3\sigma^\mu\Bigl[ i\ k_{2\lambda}\ \tilde h_\mu\ ^\lambda
(k_1-k_2)-{i\over 2}\ \sigma_3\ \tilde\omega_\mu (k_1-k_2)-\Bigr.\nonumber\\
&&\qquad \Bigl. -i\ \tilde A_\mu (k_1-k_2)+{i\over 2} \ (k_1-k_2)_\mu\ \tilde h
(k_1-k_2)\Bigr] v_+(k_2)
\end{eqnarray*}
Substituting these matrix elements in (6.8) and letting $\Omega\to\infty$ after
some straightforward algebra one obtains
\setcounter{equation}{0}
\renewcommand{\theequation}{6.10\alph{equation}}
\begin{eqnarray}
A_L & = & \vert\Lambda\vert\int {d^2p\over (2\pi )^2}\ e^{ip\cdot x}\
G_{\mu\nu}(p)\ \tilde h_{\mu\nu}(p)\\
A_{Maxw.} & = & \vert\Lambda\vert\int {d^2p\over (2\pi )^2}\ e^{ip\cdot x}\
G(p)\ i\ \varepsilon_{\mu\nu}\ p_\nu\ \tilde A_\mu (p)\\
T_\mu & = & \vert\Lambda\vert\int {d^2p\over (2\pi )^2}\ e^{ip\cdot x}\
G_{\mu\lambda}(p)\ i\ \varepsilon_{\sigma\nu}\ p_\nu\ \tilde h_{\sigma\lambda}
(p)
\end{eqnarray}
where
\setcounter{equation}{0}
\renewcommand{\theequation}{6.11\alph{equation}}
\begin{eqnarray}
G_{\mu\nu} (p) & = & \int {d^2k\over (2\pi )^2}\ {k_\mu\ k_\nu\over
\left(\omega\left( k+{p\over 2}\right) +\omega \left( k-{p\over
2}\right)\right)\omega\left( k+{p\over 2}\right)\omega\left( k-{p\over
2}\right)}\\ G(p) & = &\int {d^2k\over (2\pi )^2}\ {1\over \left(\omega \left(
k+{p\over 2}\right) +\omega\left( k-{p\over 2}\right)\right)\omega \left(
k+{p\over 2}\right)\omega\left( k-{p\over 2}\right)}
\end{eqnarray}
and the $\vert\Lambda\vert\to\infty$ limit of (6.10) is understood. The kernel
$G(p)$ is given by a convergent integral. Therefore in the limit of
$\vert\Lambda\vert\to\infty$ it becomes $G(0)={1\over 4\pi\vert\Lambda\vert}$.
Inserting this in (6.10b) we obtain
\setcounter{equation}{11}
\renewcommand{\theequation}{6.\arabic{equation}}
\begin{equation}
A_{Maxw.}={1\over 4\pi}\ \varepsilon_{\mu\nu}\ \partial_\nu\ A_\mu (x)
\end{equation}
To evaluate the $\vert\Lambda\vert\to\infty$ limit of $G_{\mu\nu}(p)$, we
expand the integrand in powers of $\displaystyle{p\over\vert\Lambda\vert}$. The
leading term will be a $p$--independent linearly divergent integral which
should
be handled by adopting a suitable subtraction scheme. The terms of order
$\displaystyle{1\over\vert\Lambda\vert}$, on the other hand, are finite and are
given by
\begin{equation}
G_{\mu\nu}(p) ={1\over 48\pi}\ (p_\mu\ p_\nu -p^2\ \delta_{\mu\nu})\
{1\over\vert\Lambda\vert}+0\left({1\over\vert\Lambda\vert^3}\right)
\end{equation}
Substituting this in (6.10a) and (6.10c) we obtain
\setcounter{equation}{0}
\renewcommand{\theequation}{6.14\alph{equation}}
\begin{eqnarray}
A_L & = & {1\over 48\pi}\ (\partial^2\ \delta_{\mu\nu} -\partial_\mu\
\partial_\nu ) h_{\mu\nu}(x)\\
T_\mu & = & {1\over 48\pi}\
(\partial^2\delta_{\mu\lambda}-\partial_\mu\partial_\lambda )
\varepsilon_{\sigma\nu}\ \partial_\nu\ h_{\sigma\lambda}(x)
\end{eqnarray}
These results which were contained in [14] agree with the linearized
versions of the corresponding equations of [15].

\section{Outlook}

In our opinion one of the more urgent questions to be addressed in the overlap
approach is to extend the calculations of Sec.5 to all orders in $A$ in a
properly regularized theory, i.e. on the lattice. This will increase our
confidence in the validity of the approach. Of course, eventually, the dynamics
of the boson fields should also be incorporated.

Ultimately the value of the scheme will depend on its usefullness in
non--perturbative studies of standard model.

\newpage

\noindent{\Large \bf Appendix:\quad The $\Lambda$--discontinuity}
\bigskip

The discontinuity at $\Lambda =0$ of the functional $\Phi (\theta ,A)$ is
computed for the 2--dimensional case. This concerns the first order term
obtained in Sec.4
\setcounter{equation}{0}
\renewcommand{\theequation}{A.\arabic{equation}}
\begin{equation}
\Phi_\pm (\theta ,A)=\int\left({dp\over 2\pi}\right)^2\ \widetilde\theta_\alpha
(-p)\ F^\mu_\pm (p)\ \widetilde A^\alpha_\mu (p)
\end{equation}
where $F^\mu_+(p)$ is given by (4.12). Of special concern is its first
derivative evaluated at $p=0$, given by (4.13) and (4.14),
\begin{equation}
F^{\mu ,\nu}_\pm (0)  =  {i\over 4}\
\varepsilon^{\mu\nu}\int_{BZ}\left({dk\over 2\pi}\right)^2\
{1\over\omega^3_\pm}\ \langle T_c\rangle\Bigl\{ (B\pm\vert\Lambda\vert )\det C
+\Bigr.
 \Bigl. \left( C_\alpha\ ^{,\beta}\ B^{,\alpha} -C_\alpha\ ^{,\alpha}\
B^{,\beta}\right) C_\beta\Bigr\}
\end{equation}
where $\langle T_c\rangle$ is defined by the flavour trace,
\begin{eqnarray}
{\rm tr} (T_c\ \theta\ A_\mu ) & = & {\rm tr} (T_c\ T_\beta\ T_\alpha )\
\theta^\beta\ A^\alpha_\mu\nonumber \\
& = & \langle T_c\rangle\ \theta_\alpha\ A^\alpha_\mu
\end{eqnarray}
In the integral $B$ and $C_\mu$ are functions of $k$ and $\det C$ is the
determinant of $\partial C_\mu /\partial k_\nu$. The energies are given by
\begin{equation}
\omega_\pm =\sqrt{g^{\mu\nu}\ C_\mu\ C_\nu +(B\pm\vert\Lambda\vert )^2}
\end{equation}
where $g_{\mu\nu}$ is the lattice metric normalized such that $\det g=1$. The
integral can develop infra--red singularities for $\Lambda =0$ at points where
$\omega_\pm$ vanishes, i.e. wherever $B=C_\mu =0$. These singularities
contribute to the discontinuity, $F_+-F_-$, at $\Lambda =0$.

The purpose of this appendix is to estimate the contribution of one such
singularity. Suppose that $C_\mu$ and $B$ have simple zeroes at $k=\hat k$,
\begin{eqnarray}
C_\mu (k) & = & C_\mu\ ^{,\nu}(\hat k)\ (k-\hat k)_\nu +\dots \equiv p_\mu
+\dots\nonumber \\
B(k) & = & B^{,\nu}(\hat k)\ (k-\hat k)_\nu +\dots \equiv b^\mu\ p_\mu +\dots
\end{eqnarray}
To estimate the small $\Lambda$ behaviour we can use the method of steepest
descents. Write
$$
{1\over\omega^3}={1\over\Gamma (3/2)}\int^\infty_0d\alpha\ \alpha^{1/2}\
e^{-\alpha\omega^2}
$$
and express (A.2) as a Laplace transform
\begin{equation}
F^{\mu ,\nu}_\pm (0) = i\ \varepsilon^{\mu\nu}\ \langle T_c\rangle\int^\infty_0
d\alpha\ e^{-\alpha\Lambda^2}\ K_\pm (\alpha ,\Lambda )
\end{equation}
where
\begin{equation}
K_\pm (\alpha ,\Lambda )  =  {\alpha^{1/2}\over 4\Gamma
(3/2)}\int\left({dk\over 2\pi}\right)^2\ e^{-\alpha
(\omega^2_\pm-\Lambda^2)} \left\{ (B\pm\vert\Lambda\vert )\det C
+(C_\alpha\ ^{,\beta}\ B^{,\alpha}-C_\alpha\ ^{,\alpha}\
B^{,\beta})C_\beta\right\}
\end{equation}
The small $\Lambda$ behaviour of (A.6) is determined by the large $\alpha$
behaviour of (A.7) and this is dominated by contributions from the
neighbourhood of $\hat k$, i.e. small values of the momentum $p_\mu$ defined in
(A.5). Change the integration variable from $k$ to $p$ and expand in powers of
$p$ so that, for large $\alpha$,
\begin{eqnarray}
K_\pm (\alpha ,\Lambda ) &\simeq& {\alpha^{1/2}\over 4\Gamma (3/2)}\ {\det
C\over\vert\det C\vert}\int\Bigl({dp\over 2\pi}\Bigr)^2\exp\Biggl[ -\alpha
\Bigl(\pm 2\vert\Lambda\vert b^\mu p_\mu +(g^{\mu\nu} +b^\mu b^\nu )p_\mu p_\nu
+\dots \Bigr)\Biggr]\cdot\nonumber \\
&&\hspace{4cm}\cdot\left\{ \pm\vert\Lambda\vert +\left( b^\beta +{C_\alpha\
^{,\beta}\ B^{,\alpha}-C_\alpha\ ^{,\alpha}\ B^{,\beta}\over \det C}\right)
p_\beta +\dots\right\}\nonumber \\
&&
\end{eqnarray}
The integration can be extended to the entire plane without affecting the
asymptotic development in inverse powers of $\alpha$. Use the Gaussian formula
\begin{equation}
\int\left({dp\over 2\pi}\right)^2\ e^{-\alpha M^{\mu\nu} p_\mu p_\nu+\xi^\mu
p_\mu}={1\over 4\pi\alpha}\ \vert\det M\vert^{-1/2}\exp\left[{1\over
4\alpha}\xi^\mu M^{-1}_{\mu\nu} \xi^\nu\right]
\end{equation}
Here we have
\begin{eqnarray*}
M^{\mu\nu} & = & g^{\mu\nu} +b^\mu b^\nu\\
\xi^\mu & = & \mp 2\alpha\vert\Lambda\vert b^\mu +\eta^\mu
\end{eqnarray*}
where $\eta^\mu$ is small. Since
$$
\det M =1+b^2\quad {\rm and}\quad M^{-1}_{\mu\nu} =g_{\mu\nu}-(1+b^2)^{-1}\
b_\mu b_\nu
$$
the right--hand side of (A.9) is given by
$$
{1\over 4\pi\alpha}\ (1+b^2)^{-1/2}\exp\left({\alpha\Lambda^2b^2\over
1+b^2}\right)\cdot\left( 1\mp {\vert\Lambda\vert b_\mu\eta^\mu\over
1+b^2}+0(\eta^2)\right)
$$
Hence, (A.8) reduces to
\begin{eqnarray}
K_\pm (\alpha ,\Lambda ) &\simeq & \pm\vert\Lambda\vert\ {\det
C\over\vert\det C\vert}\ {\alpha^{-1/2}(1+b^2)^{-1/2}\over 16\pi\ \Gamma
(3/2)}\exp\left({\alpha\Lambda^2b^2\over 1+b^2}\right)\cdot\nonumber \\
&&\nonumber \\
&&\cdot\left[ 1-\left( b^\beta +{C_\alpha\ ^{,\beta}\ B^{,\alpha} -C_\alpha\
^{,\alpha}\ B^{,\beta}\over\det C}\right)\ {b_\beta\over 1+b^2}+0\left(\Lambda
,{1\over\alpha\Lambda}\right)\right]
\end{eqnarray}
The neglected terms are unimportant because, in effect,
$\alpha\sim\Lambda^{-2}$ for $\Lambda\to 0$. The second term in the square
brackets vanishes because, on using the definitions (4.5),
\begin{eqnarray*}
(C_\alpha\ ^{,\beta}\ B^{,\alpha} - C_\alpha\ ^{,\alpha}\ B^{,\beta})b_\beta &
=
& (C_\alpha\ ^{,\beta}\ C_\gamma\ ^{,\alpha} -C_\alpha\ ^{,\alpha}\ C_\gamma\
^{,\beta})b^\gamma b_\beta\nonumber \\
& = & -b^2\det C
\end{eqnarray*}
Finally, substituting (A.10) into (A.6) one obtains
$$
F^{\mu ,\nu}_\pm (0) =\pm\ i\ \varepsilon^{\mu\nu}\ \langle T_c\rangle\ {1\over
8\pi}\ {\det C\over\vert\det C\vert}+0(\Lambda )
$$
The discontinuity at $\Lambda =0$ therefore receives the contribution
\begin{equation}
F^{\mu ,\nu}_+(0)-F^{\mu ,\nu}_-(0)=i\ \varepsilon^{\mu\nu}\ \langle T_c\rangle
\ {1\over 4\pi}\ {\det C\over \vert \det C\vert}
\end{equation}
from every simple zero on the torus. Note that this result does not depend
on $B^{,\mu}(\hat k)$. The order of the zero in $B$ is not important.

\end{document}